\documentclass[11pt,a4paper]{article}
\usepackage[utf8]{inputenc}
\pdfoutput=1
\usepackage{jheppub}
\usepackage{amsfonts}
\usepackage{amsmath}
\usepackage{color}
\usepackage[normalem]{ulem}
\usepackage{verbatim}
\usepackage{rotating, graphicx}
\usepackage{subcaption}

\newcommand{\beq}{\begin{equation}}
\newcommand{\eeq}{\end{equation}}
\newcommand{\bea}{\begin{eqnarray}}
\newcommand{\eea}{\end{eqnarray}}

\newcommand{\dbar}{d\hspace*{-0.08em}\bar{}\hspace*{0.1em}}
\title{Towards Colour Flow Evolution at Two Loops}
\author{Simon Pl\"atzer and Ines Ruffa}
\affiliation{Particle Physics, Faculty of Physics,\\
  \hspace*{2ex}University of Vienna, 1090 Wien, Austria}
\affiliation{Erwin-Schr\"odinger-Institute for Mathematics and Physics,\\
  \hspace*{2ex}University of Vienna, 1090 Wien, Austria}

\emailAdd{simon.plaetzer@univie.ac.at}
\emailAdd{ines.ruffa@univie.ac.at}

\preprint{\begin{flushright}UWTHPH-2020-11\\ MCnet-20-28\end{flushright}}

\abstract{We calculate the two-loop and one-loop/one-emission
  contributions required for soft gluon evolution at the
  next-to-leading order. The colour structures are expressed in the
  colour flow basis, and the kinematic dependence and loop integrals
  are expressed in terms of multiple cuts and phase-space-like
  integrals. This directly allows to use them in the resummation of
  non-global observables and improved parton shower algorithms beyond
  the leading order and beyond the leading colour limit. Within the
  colour flow basis it becomes apparent that correlations beyond a
  dipole picture emerge even in colour-diagonal elements of the
  virtual corrections.}

\begin{document}

\maketitle
\flushbottom

\section{Introduction}
\label{sec:Introduction}

Precise predictions for collider physics experiments require to
include QCD corrections both at fixed-order and within resummed
perturbation theory. While fixed-order calculations reliably predict
inclusive quantities and the coarse spectra of jets at high transverse
momenta, infrared sensitive observables require resummed perturbation
theory in order to capture the dominant, logarithmic contributions
appearing at each order in the perturbative series. A complementary
approach is the use of multi-purpose Monte Carlo event generators,
which provide very detailed simulation of realistic final states,
including phenomenological models of how a high-multiplicity partonic
final state with small inter-parton scales converts into the observed
hadrons. The component central to event generators is the parton
shower algorithm which describes the evolution from large to small
momentum transfers and also facilitates the resummation of large
logarithmic contributions.

The resummation of non-global observables
\cite{Dasgupta:2001sh,Banfi:2002hw} requires a framework of parton
evolution at the amplitude level
\cite{Becher:2016mmh,Caron-Huot:2015bja,Martinez:2018ffw}, and such a
formalism is very promising in providing a theoretical framework which
allows to systematically construct parton shower algorithms beyond the
currently adopted approximations \cite{Nagy:2014mqa,Forshaw:2019ver},
and with the highest level of control over their accuracy. Such a
formalism has recently been adopted to make decisive statements about
the accuracy of parton showers
\cite{Forshaw:2020wrq,Holguin:2020joq,Dasgupta:2020fwr,Hamilton:2020rcu,Nagy:2020rmk}.
Solving the evolution equations which do resum non-global logarithms
requires Monte Carlo methods \cite{Balsiger:2018ezi} and has recently
been extended beyond the leading-$N$ limit
\cite{Platzer:2013fha,DeAngelis:2020rvq}, addressing a dedicated
resummation algorithm before aiming at a more versatile simulation
within an event generator. The same formalism also allows to analyze
approaches to improve existing parton showers beyond the leading-$N$
limit \cite{Platzer:2012np,Platzer:2018pmd,Hoeche:2020nsx} in order to
show which colour suppressed contributions are actually taken into
account \cite{Holguin:2020oui}, highlighting the fact that amplitude
level evolution will go beyond a probabilistic approach in the sense
that unitarity cannot be naively employed anymore.  A connection to
colour reconnection models has recently been highlighted, as well
\cite{Gieseke:2018gff}.

In the present work we address key ingredients of extending the
amplitude level evolution beyond the leading order, and towards
including soft gluon effects at second order in the strong coupling,
within the double-soft limit. The evolution at this order will require
two-loop, one-loop one-emission and double-emission diagrams to be
available in a form appropriate to be handled by a numerical
code. This will provide the entry point to include triple-collinear
and soft-collinear configurations, and serves the purpose of
clarifying the structure of the evolution in colour space similarly to
the one-loop anomalous dimension analyzed in \cite{Platzer:2013fha}.
We also establish methodology in order to systematically obtain the
virtual corrections required at this order in a representation of
phase-space type integrals, which is complementary to approaches which
perform the actual integrals and make the divergencies explicit in
terms of poles of the dimensional regularisation parameter $\epsilon$,
see {\it e.g.} results on one-, two- and three-loop divergencies
\cite{Catani:1996vz,Catani:1998bh,Becher:2009cu}, as well as the
one-loop corrections to the emission of a soft gluon
\cite{Catani:2000pi}. The resulting anomalous dimensions for soft
gluon exchange have been discussed in more detail in
\cite{Aybat:2006wq,Almelid:2015jia}, and their role in resummation of
soft gluon effects in hadronic $2\to 2$ scatterings has been pioneered
in \cite{Kidonakis:1998nf,Contopanagos:1996nh,Dokshitzer:2005ig} and
recently summarised {\it e.g.} in \cite{Kidonakis:2020gxo}. The
resummation of soft gluon corrections has also been considered for a
larger class than $2\to 2$ scatterings \cite{Sjodahl:2008fz}, and
general algorithms for dealing with the colour structure are available
in \cite{Sjodahl:2009wx}. Our approach focuses on a general algorithm
using the colour flow basis, and a formulation appropriate for an
arbitrary number of coloured legs.\footnote{Here we only consider
outgoing partons, though a generalisation of our methods to incoming
partons and colour correlations between in- and outgoing partons is
straightforward.} Within our formalism we are also in the position to
isolate the imaginary parts of the virtual corrections (which we leave
in explicit and detailed form for a future publication), as well as
the possibility to straightforwardly implement subtractions, and the
evaluation of virtual corrections in a manner inherent to parton
branching algorithms based on (weighted) Sudakov veto algorithms
\cite{Platzer:2011dq,Bellm:2016voq,Olsson:2019wvr}.

The present work is structured as follows: In Sec.~\ref{sec:Review} we
will first review basic principles of soft gluon evolution in colour
space, and in particular at the leading order using one-loop soft
exchanges and single emission Eikonal currents. We will also review
that the colour structures obtained at this level give rise to a
systematic expansion around the leading-$N$ limit. In
Sec.~\ref{sec:NLOColourStructures} we will then perform a detailed
analysis of the colour structures encountered in the two-loop, and the
one-loop one-emission cases using the colour flow basis which has
proven to be computationally advantageous; it also closely connects to
the language actual parton shower algorithms are formulated in. In
Sec.~\ref{sec:Kernels} we will then outline our approach to the
virtual contributions, and in particular we will detail the
algorithmic way of casting the virtual corrections into
phase-space-type integrals by means of the Feynman tree theorem
\cite{Feynman:1963ax,Catani:2008xa}, which we generalise to account
for Eikonal propagators, as well as complications appearing at two
loops such as propagators raised to higher powers. Detailed formulae
of our results are collected in several appendices and will serve as
input to future work which we discuss in a summary and outlook in
Sec.~\ref{sec:Outlook}.

\section{Soft gluon evolution and the colour flow basis}
\label{sec:Review}

Amplitude evolution algorithms like the approach outlined in
\cite{Forshaw:2019ver}, and the resummation of non-global logarithms
\cite{Becher:2016mmh}, proceed through evolution equations in colour
space which govern the contribution to the cross section originating
from $n$ hard partons. They generically are expected to be of the form
\begin{equation}
  \label{eqs:evol}
  E\frac{\partial}{\partial E} {\mathbf A}_n(E) = 
  {\mathbf \Gamma}_n(E) {\mathbf A}_n(E) + {\mathbf A}_n(E) {\mathbf \Gamma}^\dagger_n(E)
 - \sum_{k} {\mathbf R}_{n}^{(k)}(E)  {\mathbf A}_{n-k}(E) {\mathbf R}_{n}^{(k),\dagger}(E) \ ,
\end{equation}
while the final cross section is a trace over the hard function
${\mathbf A}_n$ and a soft function ${\mathbf S}_n$, which determine
the final state configurations on which the observable can then be
evaluated,
\begin{equation}
  \sigma[u] = \sum_n \int {\rm Tr}\left[{\mathbf A}_n {\mathbf
      S}_n\right] u(q_1,...,q_n) {\rm d}\phi(q_1,...,q_n|Q) \ .
\end{equation}
Here $q_1,...,q_n$ denote the final state particles and ${\rm d}\phi$
is the associated phase space measure which can be approximated for
soft emissions for which momentum conservation is a sub-leading
effect. The evolution equation (\ref{eqs:evol}) comprises the soft
anomalous dimension matrices ${\mathbf \Gamma}_n$ as well as emission
operators ${\mathbf R}_{n}^{(k)}$ which describe how $k$ partons are
emitted from an $n-k$ parton state. Both ${\mathbf \Gamma}$ and
${\mathbf R}$ have perturbative expansions and originate after
diagrammatic recursions in the soft limit have appropriately been
subtracted and renormalised. Upon taking matrix elements in between
different colour flows $\tau,\sigma$
\cite{Platzer:2013fha,Martinez:2018ffw}, the virtual corrections can
be expressed as an expansion in the 't Hooft coupling
\begin{equation}
  [\tau|{\mathbf \Gamma}|\sigma\rangle = (\alpha_s N) [\tau|{\mathbf \Gamma}^{(1)}|\sigma\rangle
      + (\alpha_s N)^2 [\tau|{\mathbf \Gamma}^{(2)}|\sigma\rangle + ... \ .
\label{eqs:thooftexpansion}
\end{equation}
For a jet cross section at leading order ${\mathbf S}_n$ is the
identity operator in colour space, and the evolution of ${\mathbf
  A}_n$ is driven by iterating single, soft emissions and the one-loop
soft anomalous dimension, which results from a combination of the real
emission and virtual contributions subject to the resolution scale
$E$. To be more precise, we allow to emit gluons above the evolution
scale $E$, and combine the contribution from lower energies to cancel
the infrared divergencies in the one-loop integral. The remaining
(artificial) ultraviolet divergences in the one-loop and one-emission
cross section stemming from the soft gluon approximation are then
absorbed to all orders into a renormalisation of the hard and soft
functions, giving rise to the evolution equation stated in
Eq.~(\ref{eqs:evol}). More details of such an approach will be
discussed in upcoming work.

The purpose of the present work is to investigate in detail the
structure of the Feynman diagrams contributing to the evolution at the
next-to-leading order in colour space, as well as to explore
strategies how these contributions can be manipulated to allow for the
appropriate subtractions at the level of phase-space-type
integrals. This will enable us to address the observable dependence in
a most differential way. The basis-independent colour space formalism
\cite{Catani:1996vz} is very well suited for general investigations of
the colour structure of soft gluon contributions, however in a
practical implementation one has to choose a basis and express the
amplitudes and colour space operators as complex vectors and matrices
with respect to the basis chosen.

As the dimensionality of the colour space grows asymptotically as a
factorial with the number of external legs, the handling of these
objects soon becomes intractable and a numerical solver of evolution
equations of the type of Eq.~(\ref{eqs:evol}) can only proceed by using
Monte Carlo methods to sample over the different colour structures
involved, possibly paired with approximation methods to keep track of
certain classes of $1/N$ suppressed contributions to the cross section
\cite{Platzer:2013fha,Martinez:2018ffw}. From a computational point of
view, the colour flow basis has proven to be useful in such an
approach, and a practical code has been based on it
\cite{DeAngelis:2020rvq}. In order to be able to solve the evolution
equations in this approach at the next order, we choose to express the
anomalous dimension and emission matrices not only in the basis
independent notation, but also in the colour flow basis. To make the
connection between the two clear we will review the one-loop case in
the next section, and we will then outline our approach to the loop
integrals involved.

\subsection{Leading-order Evolution}

In the leading order evolution, the one-loop contribution takes the
form
\begin{equation}
\bold{\Gamma}^{(1)}=\frac{1}{2} \sum_{i,j} \Omega^{(1)}_{ij}\ \frac{1}{N}\bold{T}_i \cdot \bold{T}_j \ ,
\end{equation}
where the individual coefficients $\Omega^{(1)}_{ij}$ can be
deduced by casting the loop integral corresponding to an Eikonal
exchange of a gluon in between two (in this case, outgoing) external
lines $i$ and $j$ into the form
\begin{equation}
  \Omega^{(1)}_{ij}\ =i \mu^{2\epsilon}
  \int \frac{{\rm \dbar}^dk}{i\pi^{d/2}} \frac{p_i\cdot p_j}{(k^2+i 0)(p_i\cdot k+i 0)(p_j\cdot k-i 0)} =
\int_0^\infty \frac{{\rm d}E}{E} \left(\frac{\mu^2}{E^2}\right)^\epsilon \omega^{(ij)} \ ,
\label{eqs:gammadefinition}
\end{equation}
where we have defined $\frac{\mathrm{\dbar}^dk}{i\pi^{d/2}}\equiv 4\pi \frac{\mathrm{d}^dk}{(2\pi)^d}$.
After combining the virtual contributions with the real emission the
soft singularity $E\to 0$ will cancel and we are left with an
ultraviolet divergence in $E$ only (collinear divergences in
$\omega^{(ij)}$ will cancel with the real emission when calculating a
full cross section).  At leading order, $\omega^{(ij)}$ is independent
of the scales $\mu$ and $E$, and only depends on the hard parton's
directions of motion rather than their energies. We find in the
case of two outgoing or two incoming lines that
\begin{equation}
  \omega^{(ij)} = \frac{(2\pi)^{2\epsilon}}{\pi}\left[\int \frac{{\rm d}\Omega^{(d-2)}}{4\pi}
  \frac{n_i\cdot n_j}{n_i\cdot n\ n\cdot n_j} - i\pi \int \frac{{\rm d}\Omega^{(d-3)}}{2\pi}\right] \ .
\end{equation}
In this case, we have assumed an ordering in energy, though other
ordering variables are possible and will give rise to different forms of
the anomalous dimension; we therefore discuss the loop integrals
without making a reference to a particular ordering variable or
observable. Explicit results for energy and $p_\perp$ ordering will be
presented in an upcoming publication.

In the colour flow basis, the same quantity takes the form
\begin{equation}
[\tau|{\mathbf \Gamma}^{(1)}|\sigma\rangle =  \left(\Gamma_\sigma^{(1)}+\frac{1}{N^2}\rho^{(1)}\right) \delta_{\sigma\tau} + \frac{1}{N} \Sigma_{\sigma\tau}^{(1)}\ ,
\end{equation}
which is readily verified by using the decomposition of the colour
charge correlators in the colour flow basis
\begin{equation}
\begin{split}
[\tau| \bold{T}_i\cdot \bold{T}_j |\sigma \rangle= -N \delta_{\tau \sigma} \left[\lambda_i \bar{\lambda}_j \delta_{c_i, \sigma^{-1}(\bar{c}_j)} +\bar{\lambda}_i \lambda_j \delta_{c_j, \sigma^{-1}(\bar{c}_i)} +\frac{1}{N^2} (\lambda_i-\bar{\lambda}_i) (\lambda_j-\bar{\lambda}_j)\right] \\
+ \sum_{(ab)} \delta_{\tau_{(ab)},\sigma} \left(\lambda_i \lambda_j \delta_{(ab),(c_i c_j)}+ \bar{\lambda}_i \bar{\lambda}_j \delta_{(ab),(\sigma^{-1}(\bar{c}_i) \sigma^{-1}(\bar{c}_j))} \right. \\
\left. -\lambda_i \bar{\lambda}_j \delta_{(ab),(c_i \sigma^{-1}(\bar{c}_j))} - \bar{\lambda}_i \lambda_j \delta_{(ab),(c_j \sigma^{-1}(\bar{c}_i))}\right) \ .
\end{split}
\label{eqws:1loopcorrelator}
\end{equation}
In fact, we define the coefficients $\Gamma_\sigma^{(1)}$,
$\rho^{(1)}$ and $\Sigma_{\tau\sigma}^{(1)}$ through this relation;
implementations performing this calculation are available from the
authors.  We shall obtain similar identities for the colour structures
required for the evolution in the next order. At this point it is
important to remark that we can explicitly identify what the leading,
colour diagonal contributions are, and how the very sparse elements in
the off-diagonal part of the anomalous dimension matrix can be
addressed efficiently. This knowledge allows for an efficient Monte
Carlo in colour space as well as a systematic expansion around the
large-$N$ limit mentioned earlier.

While, in a perturbative expansion, one would treat the diagonal,
$1/N^2$ suppressed bit as a correction this turns out not to be a
viable approach in the presence of collinear contributions: in this
case, dropping the $1/N^2$ contribution amounts to effectively
replacing $C_F$ by $C_A/2$ in the quark splitting function, which
would thus not properly take into account logarithmic contributions of
soft- and hard-collinear origin. For this reason we stress that an
appropriate expansion around the large-$N$ limit would actually need
to be seen as an expansion around the colour diagonal part (also
referred to as $d'$ approximations in \cite{DeAngelis:2020rvq}),
something which will be of great importance when discussing the colour
structures appearing the two-loop case. Also note that the $1/N^2$
suppression in the one-loop anomalous dimension might possibly be
overcome since $\rho$ contains a sum over all pairs of quarks and
antiquarks, not only colour connected dipoles.

\section{Colour structures at next-to-leading order}
\label{sec:NLOColourStructures}

In order to facilitate the evolution at the next order, several
ingredients are required. In this paper we concentrate on the
evolution of the hard function ${\mathbf A}_n$, for which we require
knowledge of the two-loop, one-loop and one emission, and two-emission
soft gluon contributions to a general hard process characterised by
outgoing momenta $p_i$. In this section we will discuss the colour
structures of the two loop virtual corrections, as well as the
one-loop/one-emission corrections, and how they translate into the
colour flow basis. The actual kinematic dependence and loop integrals
will be discussed in the next section, and we are not limiting
ourselves to the case of the soft limit only, such that our analysis
of the colour structures can be readily carried on towards the
inclusion of collinear effects as well as possibly massive external
partons, see {\it e.g.} \cite{Balsiger:2020ogy} for a recent
development.

\subsection{Two-loop contributions}

Starting from the individual diagrams, in the basis independent
notation the two-loop soft anomalous dimension takes the form
\begin{equation}
  \label{eqs:virtualbydiagram}
\begin{split}
N^2 \bold{\Gamma}^{(2)}&= \sum_{i, j} \left[\frac{1}{2}(\bold{T}_i\cdot \bold{T}_j)(\bold{T}_i\cdot \bold{T}_j) \Omega^{(2)}_{ij}+\frac{1}{2}\bold{T}^b_i \bold{T}^a_i \bold{T}^a_j \bold{T}^b_j \tilde{\Omega}^{(2)}_{ij}+ \bold{T}_i^b \bold{T}_i^a \bold{T}_i^b \bold{T}_j^a  \hat{\Omega}^{(2)}_{ij} \right] \\
&+\sum_{i,j,l} \left[(\bold{T}_i\cdot \bold{T}_l)(\bold{T}_i \cdot \bold{T}_j) \Omega^{(2)}_{ijl}+\frac{1}{2} f^{abc} \bold{T}^a_i \bold{T}^b_j \bold{T}^c_l  \hat{\Omega}^{(2)}_{ijl}\right] \\
&+\sum_{i,j} T_R \, (\bold{T}_i\cdot \bold{T}_j) \left[\frac{1}{2}\Omega^{(2)}_{ij, \, \mathrm{self-en.}} + \Omega^{(2)}_{ij,\, \mathrm{vertex-corr.}} \right] \ ,
\end{split}
\end{equation}
For the self-energy and the vertex-correction we have chosen to write
the colour factor as $T_R (\bold{T}_i \cdot \bold{T}_j)$. The Casimir
invariant $C_A$ of the gluon-bubble and the vertex-correction is
understood to be absorbed into the kinematical factor, in the same way
the number of flavours $n_f$ of the fermion-bubble is part of its
kinematical factor. In this way we can conveniently write a global
colour structure for these types of diagrams, where we choose the
normalisation $T_R=1/2$ (the colour structure for the gluon-bubble
and the vertex-correction can be written as $C_A/2=T_R \, C_A$ in our
convention). Table~\ref{tab:2lfactors} gives an overview of the
diagrams we consider and the colour structures we are extracting in
the basis independent notation.
\begin{table}
\begin{center}
\begin{tabular}{|	p{3cm}	|	p{3cm}	|	p{3cm}	|}
\hline
Coefficient & Diagram & Colour-factor \\
\hline
$\Omega^{(2)}_{ij}$ & \parbox[c]{1em}{\includegraphics[width=2.3cm]{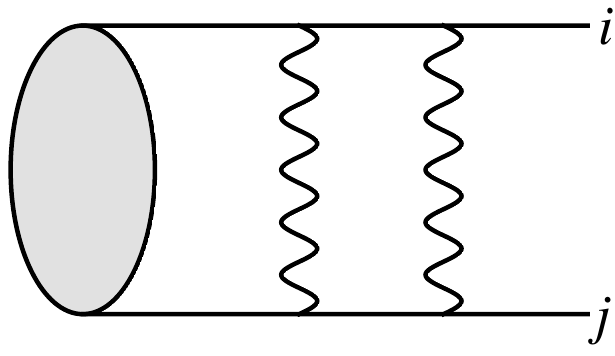}} & $(\mathbf{T}_i \cdot \mathbf{T}_j)  (\mathbf{T}_i \cdot \mathbf{T}_j)$ \\
\hline
$\tilde{\Omega}^{(2)}_{ij}$ & \parbox[c]{1em}{\includegraphics[width=2.3cm]{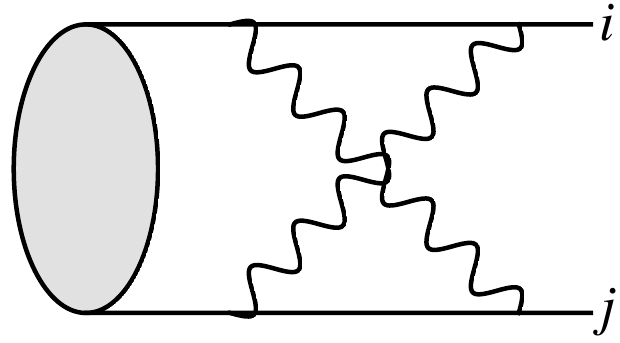}}& $\mathbf{T}_i^a \mathbf{T}_i^b \mathbf{T}_j^b \mathbf{T}_j^a$ \\
\hline
$\Omega^{(2)}_{ijl}$ & \parbox[c]{1em}{\includegraphics[width=2.3cm]{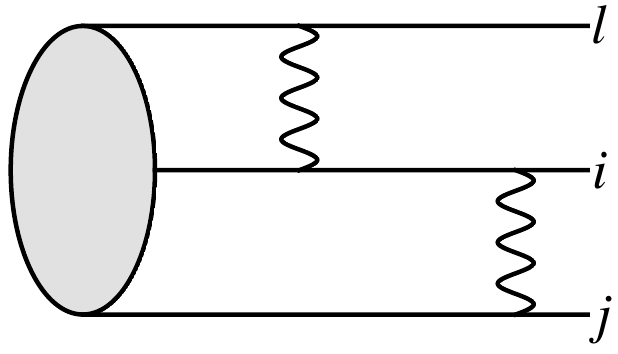}}& $(\mathbf{T}_i \cdot \mathbf{T}_l)  (\mathbf{T}_i \cdot \mathbf{T}_j)$ \\
\hline
$\hat{\Omega}^{(2)}_{ijl}$ & \parbox[c]{1em}{\includegraphics[width=2.3cm]{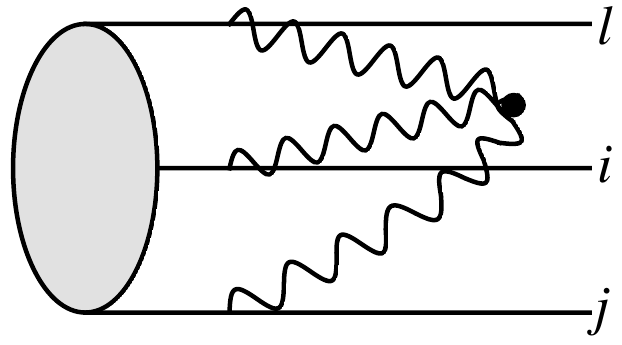}}& $if^{abc} \mathbf{T}_i^a \mathbf{T}_j^b \mathbf{T}_l^c$ \\
\hline
$\Omega^{(2)}_{ij,\mathrm{self-en.}}$ & \parbox[c]{1em}{\includegraphics[width=2.3cm]{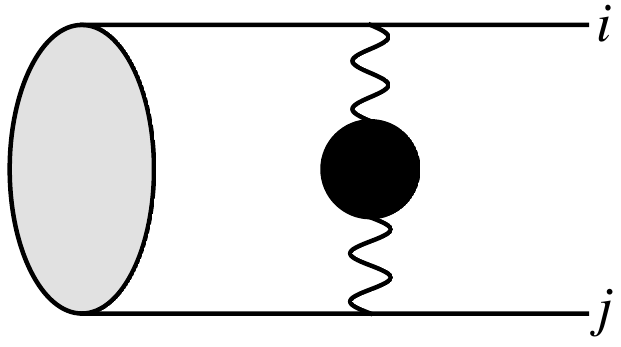}}& $T_R (\mathbf{T}_i\cdot \mathbf{T}_j)$ \\
\hline
$\Omega^{(2)}_{ij,\mathrm{vertex-corr.}}$ & \parbox[c]{1em}{\includegraphics[width=2.3cm]{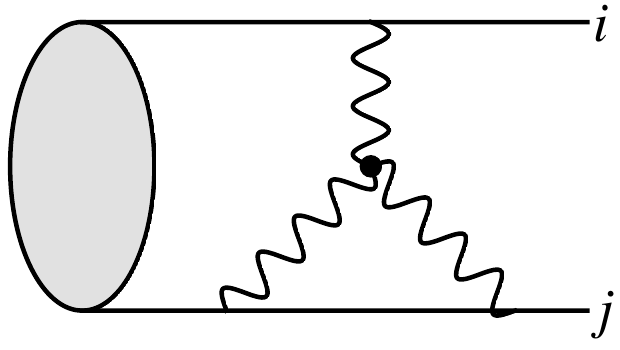}}& $T_R (\mathbf{T}_i\cdot \mathbf{T}_j)$ \\
\hline
$\hat{\Omega}^{(2)}_{ij}$ & \parbox[c]{1em}{\includegraphics[width=2.3cm]{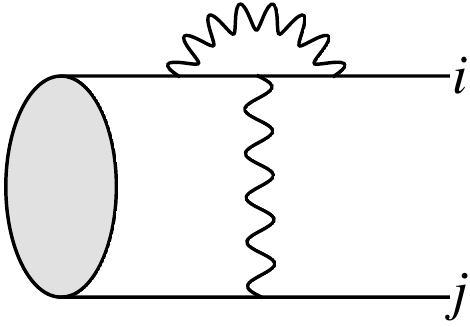}}& $\bold{T}_i^b \bold{T}_i^a \bold{T}_i^b \bold{T}_j^a$ \\
\hline

\end{tabular}
\end{center}
\caption{\label{tab:2lfactors}Two-loop diagrams contributing to the
  anomalous dimension, and the colour structures we extract in order
  to define the coefficients of the basis independent notation. See
  text for more details.}
\end{table}
Not all colour structures in this expression will in
general be independent as we can use colour conservation to
appropriately simplify the individual contributions, however
Eq.~(\ref{eqs:virtualbydiagram}) makes direct connection to the
structures encountered in individual Feynman diagrams. In fact, using
some colour algebra, we can write
\begin{equation}
 \label{eqs:virtual_colouralgebra}
\begin{split}
N^2\bold{\Gamma}^{(2)}&=\sum_{i, j} \frac{1}{2}(\bold{T}_i\cdot \bold{T}_j)(\bold{T}_i\cdot \bold{T}_j)\left[\Omega^{(2)}_{ij}+ \tilde{\Omega}^{(2)}_{ij}\right]\\
&+\sum_{i,j,l} \left[(\bold{T}_i\cdot \bold{T}_l)(\bold{T}_i\cdot \bold{T}_j) \Omega^{(2)}_{ijl}+\frac{1}{2} if^{abc} \bold{T}_i^a \bold{T}_j^b \bold{T}_l^c \hat{\Omega}^{(2)}_{ijl}\right] \\
&+ \sum_{i,j} T_R\, (\bold{T}_i\cdot \bold{T}_j) \left[\frac{1}{2}\Omega^{(2)}_{ij,\,  \mathrm{self-en.}}+ \Omega^{(2)}_{ij,\, \mathrm{vertex-corr.}}+\frac{1}{2} \tilde{\Omega}^{(2)}_{ij}+ \hat{\Omega}^{(2)}_{ij}\right] \ ,
\end{split}
\end{equation}
where we have considered unordered sums of the hard lines. As a result
we need to include factors of $1/2$ for some of the diagrams in order
to avoid overcounting. Using an unordered colour sum is convenient in
this case, since we always want to define one direction as the
emitter. Furthermore, the cutting prescription used for the loop
integrals does not necessarily show a dipole symmetry (cf. section
\ref{sec:Kernels}).  In the soft limit in Feynman gauge we can further
assume that $i\neq j$, as well as $i \neq j \neq l$ for the colour
sums, since at the two-loop level all of the diagrams involving only
one hard line are either sub-leading or vanish due to the on-shellness
of the hard lines. We point out that the part of the colour structure
pertaining to the coefficient $\hat{\Omega}_{ij}^{(2)}$ which is of
the form of the one-loop level structure after performing the colour
algebra is not anymore explicitly included in
Eq. (\ref{eqs:virtual_colouralgebra}).  \\ In view of a numerical
implementation, it is much more instructive to consider the structure
of the same quantity in a concrete basis like the colour flow basis,
and to evaluate what patterns of $1/N$ suppression arise, specifically
in a counting where the anomalous dimension loop expansion is an
expansion in the 't Hooft coupling $\alpha_s N$ (see
Eq. (\ref{eqs:thooftexpansion})), where each coefficient now admits an
expansion in $1/N$ and the different transitions the terms mediate
between colour states organised by the number of transpositions or
swaps (we refer the reader to \cite{Platzer:2013fha,Martinez:2018ffw}
for more details).

In App.~\ref{sec:ColourCorrelators} we detail how the colour
correlators appearing in the expression above can be translated into
the colour flow basis, which then gives rise to the structure
\begin{equation}
 \label{eqs:2loopcolour}
\begin{split}
  [\tau|{\mathbf \Gamma}^{(2)}|\sigma\rangle = &\left(\Gamma_\sigma^{(2)}+ \frac{1}{N^2}\left(\rho_\sigma+\tilde{\rho}\right)+\frac{1}{N^4} \rho^{(2)} \right) \delta_{\sigma\tau}  \\
&  +\frac{1}{N} \left(\Sigma_{\sigma\tau}^{(2)}+\hat{\Sigma}^{(2)}_{\sigma\tau}\right) + \frac{1}{N^3} \tilde{\Sigma}^{(2)}_{\sigma\tau}+\frac{1}{N^2} \left(\Sigma^{\prime(2)}_{\sigma\tau}+ \Sigma^{\prime\prime (2)}_{\sigma\tau} \right) \ , 
\end{split}
\end{equation}
where $\rho_\sigma$ and $\tilde{\rho}$ are additional diagonal parts,
$\hat{\Sigma}^{(2)}_{\sigma\tau}$ denotes off-diagonal parts with
single swaps, $\Sigma^{\prime(2)}_{\sigma\tau}$ and
$\Sigma^{\prime\prime(2)}_{\sigma\tau}$ represent terms with double
swaps of the form $(ab)(bc)$ and two independent single swaps
$(ab)(cd)$, respectively (cf. App.~\ref{sec:ColourCorrelators} for our
notation).

In the following we provide some examples of colour flows and identify
to which of the coefficients in Eq. (\ref{eqs:2loopcolour}) they
contribute. As with the one-loop case an implementation of this
decomposition is available from the authors. The colour-flow diagrams
are always depicted with the basis permutation $\sigma$ separated from
the rest of the diagram by the dashed line and the (anti-)colour
labels are explicitly written for the hard lines. The translation of
the diagram to the Kronecker deltas which compare the permutations
$\sigma$ and $\tau$, i.e. a translation to the corresponding parts of
the matrix elements of the colour correlators, is given.

An example of a contribution to the matrix element
$[\tau|(\bold{T}_i\cdot \bold{T}_j)(\bold{T}_i\cdot
  \bold{T}_j)|\sigma\rangle$ (cf. Eq. (\ref{A2.doubleexchange})) is
  given by the colour-flow diagram
\begin{align}
\begin{gathered}
\vspace{-0.2cm}
\includegraphics[width=4.8cm]{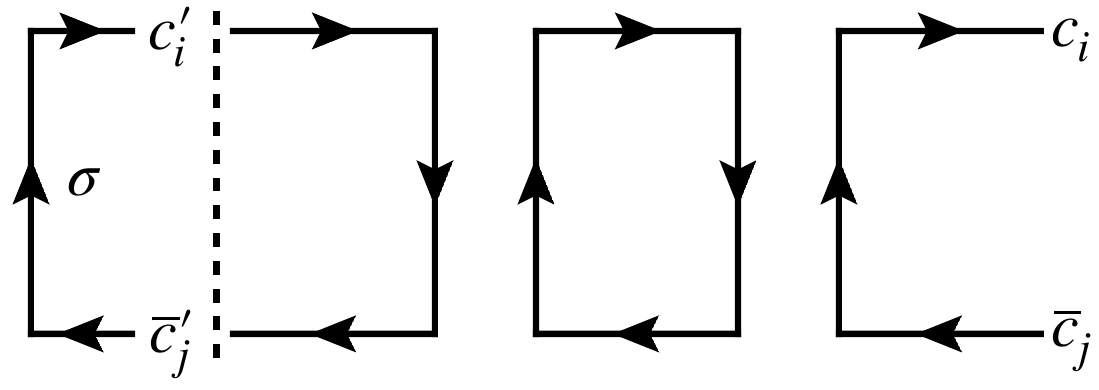}
\end{gathered}
\rightarrow N^2 \delta_{\sigma\tau} \delta_{c_i\sigma^{-1}(\overline{c}_j)} \ , &&
\end{align}
due to the colour connection in $\sigma$ this is enhanced by a factor
of $N^2$.  The diagonal structure $\rho_\sigma$ contains a
three-parton correlation from the Feynman diagram involving three hard
lines and a triple gluon vertex, it gives a colour flow of
\begin{align}
\begin{gathered}
\vspace{-0.2cm}
\includegraphics[width=4.2cm]{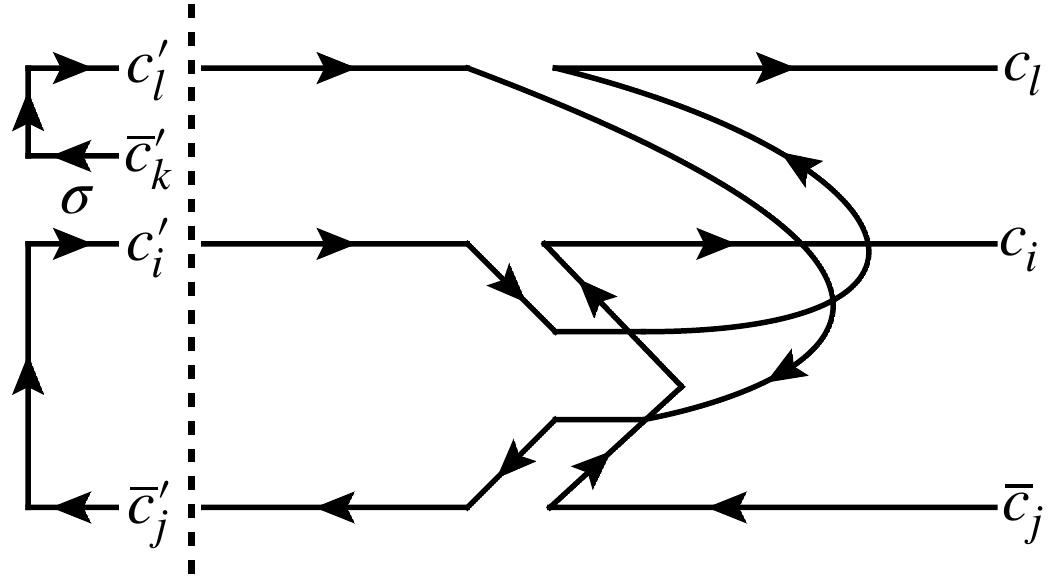}
\end{gathered}
\rightarrow \lambda_i \bar{\lambda}_j \lambda_l \delta_{\sigma\tau} \delta_{c_i\sigma^{-1}(\overline{c}_j)} \ , &&
\end{align}
and it is part of the matrix element $[\tau|\bold{T}_g \bold{T}_i
  \bold{T}_j \bold{T}_l|\sigma\rangle$
  (cf. Eq. (\ref{A2.3legtriple})).  For the gluon vertex we have
  defined that $\bold{T}_g^{abc}\equiv if^{abc}$.  For an example of
  the coefficient $\hat{\Sigma}^{(2)}_{\sigma\tau}$, which has a
  colour connection in $\sigma$ and where a single swap of colour
  labels has to be performed in order for the permutations $\sigma$
  and $\tau$ to match, consider
\begin{align}
\begin{gathered}
\vspace{-0.2cm}
\includegraphics[width=4.2cm]{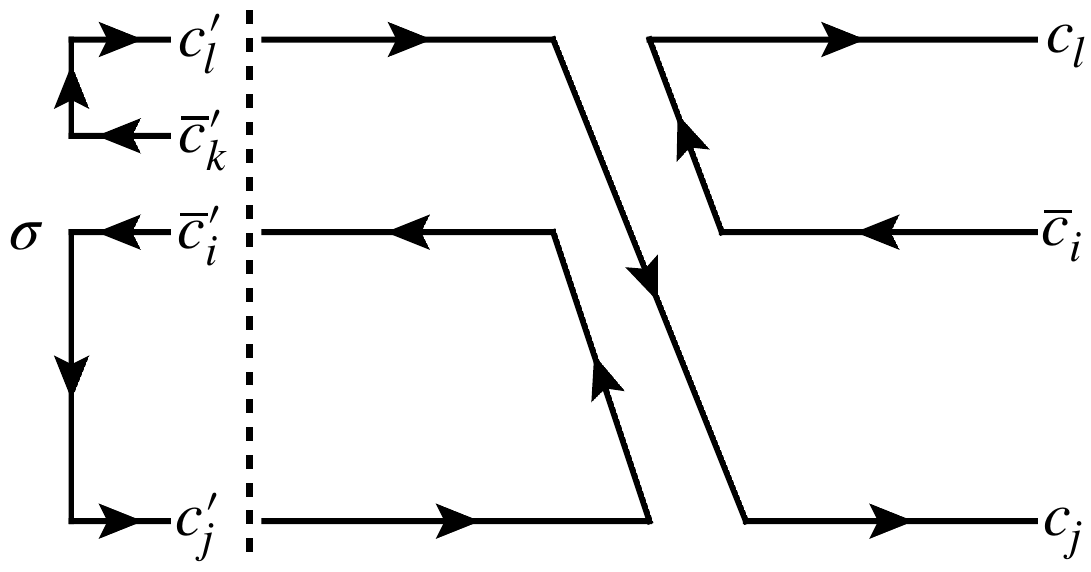}
\end{gathered}
\rightarrow N  \delta_{c_j\sigma^{-1}(\overline{c}_i)} \delta_{\sigma\tau_{(a,b)}} \delta_{(a,b)(c_j,c_l)} \ , &&
\end{align}
where this colour flow pertains to the matrix element
$[\tau|\bold{T}_g \bold{T}_i \bold{T}_j \bold{T}_l|\sigma\rangle$ as
  well.  The double swap coefficients
  $\Sigma^{\prime\prime(2)}_{\sigma\tau}$ can be exemplified by
\begin{align}
\begin{gathered}
\vspace{-0.2cm}
\includegraphics[width=4.2cm]{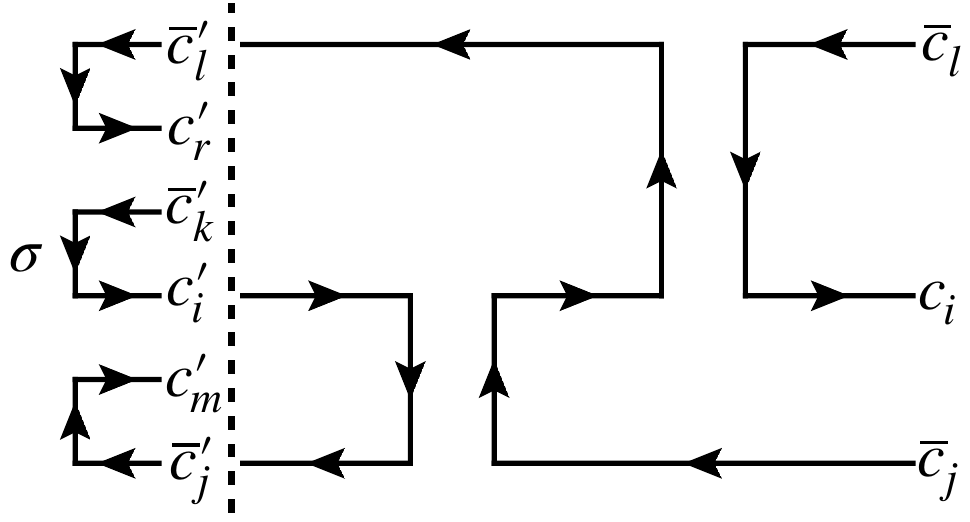}
\end{gathered}
\rightarrow \delta_{\sigma\tau_{(a,b)(b,c)}} \delta_{(a,b)(c_i,\sigma^{-1}(\overline{c}_j))}\delta_{(b,c)(\sigma^{-1}(\overline{c}_j),\sigma^{-1}(\overline{c}_l))} \ , &&
\end{align}
in this case a colour label has to be swapped twice such that the
permutations $\sigma$ and $\tau$ match. This colour flow is part of
the matrix element $[\tau|(\bold{T}_i \cdot \bold{T}_l) (\bold{T}_i
  \cdot \bold{T}_j) |\sigma \rangle$ (cf. Eq. (\ref{A2.3leg2})).

\subsection{One-loop, one-emission contributions}

Similarly to the two-loop contribution we can analyze the one-loop one
emission contributions at the level of the amplitude. Notice that at
the level of the cross section the structures we encounter will be
similar to the two-loop contribution {\it upon a fixed order
  expansion}, however in a practical evolution at the amplitude level,
the entire complexity of the one-loop correction to the emission
operator needs to be taken into account
separately. Table~\ref{tab:1l1efactors} summarises our conventions on
extracting the colour structures from the one-loop/one-emission
contributions, similarly to what we have been doing in the two-loop
case.
\begin{table}
\begin{center}
\begin{tabular}{|	p{3cm}	|	p{3cm}	|	p{3cm}	|}
\hline
Coefficient & Diagram & Colour-factor \\
\hline
$\Omega^{(1,1)}_{ij}$ & \parbox[c]{1em}{\includegraphics[width=2cm]{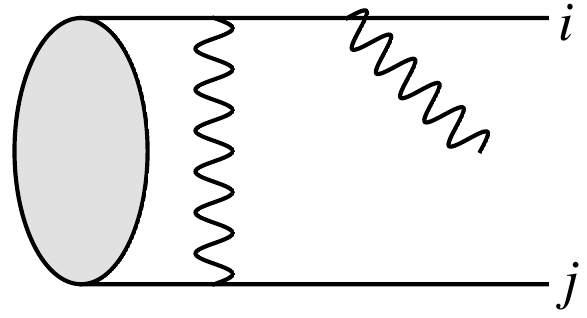}} & $\mathbf{T}_i^a (\mathbf{T}_i \cdot \mathbf{T}_j)$ \\
\hline
$\tilde{\Omega}^{(1,1)}_{ij}$ & \parbox[c]{1em}{\includegraphics[width=2cm]{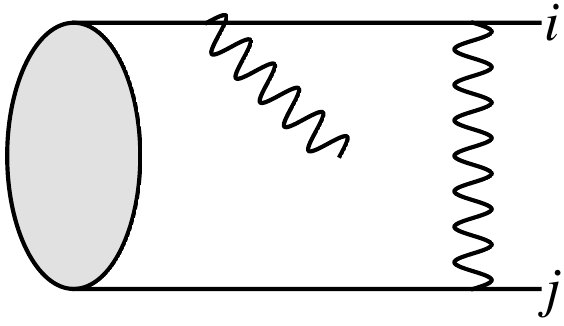}}& $(\mathbf{T}_i \cdot \mathbf{T}_j)\mathbf{T}_i^a$ \\
\hline
$\overline{\Omega}^{(1,1)}_{ij}$ & \parbox[c]{1em}{\includegraphics[width=2cm]{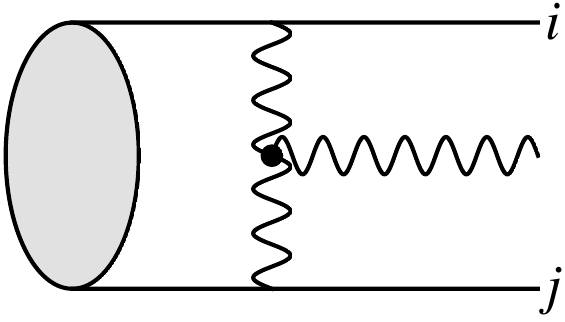}}& $i f^{abc} \mathbf{T}_i^b \mathbf{T}_j^c$ \\
\hline
$\Omega^{(1,1)}_{ijl}$ & \parbox[c]{1em}{\includegraphics[width=2cm]{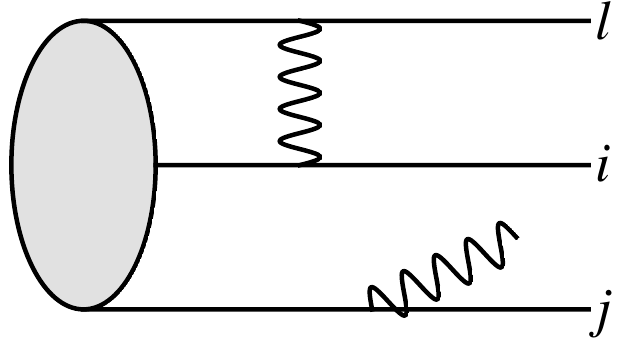}}& $\mathbf{T}_l^a (\mathbf{T}_i \cdot \mathbf{T}_j)$ \\
\hline
$\Omega^{(1,1)}_{i, \mathrm{self-en.}}$ & \parbox[c]{1em}{\includegraphics[width=2cm]{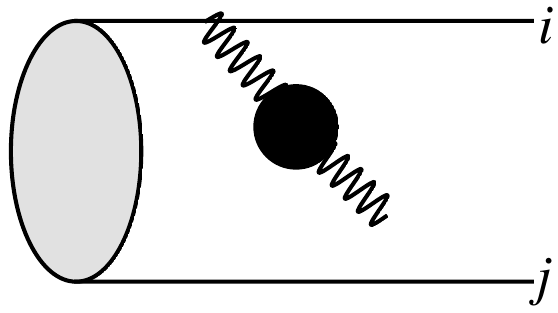}}& $T_R \mathbf{T}_i^a$ \\
\hline
$\Omega^{(1,1)}_{i,\mathrm{vertex-corr.}}$ & \parbox[c]{1em}{\includegraphics[width=2cm]{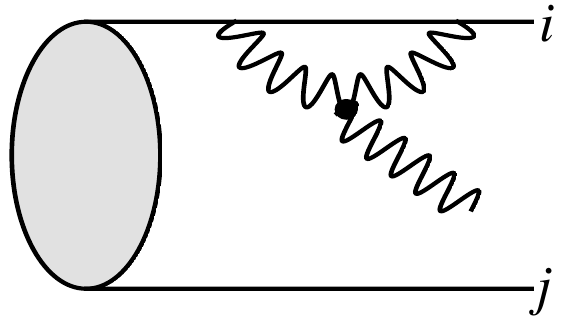}}& $T_R \mathbf{T}_i^a$ \\
\hline
$\hat{\Omega}^{(1,1)}_{ij}$ & \parbox[c]{1em}{\includegraphics[width=2.1cm]{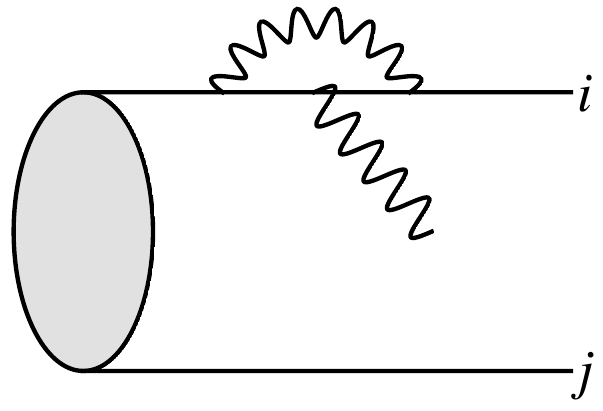}}& $\bold{T}_i^b \bold{T}_i^a \bold{T}_i^b$ \\
\hline
\end{tabular}
\end{center}
\caption{\label{tab:1l1efactors}One-loop, one-emission diagram coefficients.}
\end{table}
In the basis independent notation, the one-loop, one-emission contribution reads
\begin{equation}
\begin{split}
N^2\bold{\Gamma}^{(1,1)}&=\sum_{i,j} \left[\bold{T}^a_i (\bold{T}_i \cdot \bold{T}_j) \Omega^{(1,1)}_{ij}+(\bold{T}_i \cdot \bold{T}_j) \bold{T}^a_i \tilde{\Omega}^{(1,1)}_{ij} +\frac{1}{2} if^{abc} \bold{T}^b_i \bold{T}^c_j  \overline{\Omega}^{(1,1)}_{ij} \right] \\
&+ \sum_{i,j,l} \frac{1}{2} \bold{T}^a_l (\bold{T}_i \cdot \bold{T}_j) \Omega^{(1,1)}_{ijl}+ \sum_{i} T_R \, \bold{T}^a_i \left[\Omega^{(1,1)}_{i,\mathrm{self-en.}} + \Omega^{(1,1)}_{i, \mathrm{vertex-corr.}}+\hat{\Omega}^{(1,1)}_{ij} \right],
\end{split}
\end{equation}
and in the colour-flow basis its action is encoded in
\begin{equation}
\begin{split}
[\tau|{\mathbf \Gamma}^{(1,1)}|\sigma\rangle&= \left(\frac{1}{N^2} \rho_{\sigma\tau}+\frac{1}{N^4} \rho_\tau \right) \delta_{\sigma\tau \backslash n} \\
& +\frac{1}{N} \hat{\Sigma}^{(1,1)}_{\sigma\tau}+ \frac{1}{N^3}\left(\tilde{\Sigma}^{(1,1)}_{\sigma\tau}+ \hat{\tilde{\Sigma}}_{\sigma\tau}\right)+ \frac{1}{N^2} \left(\Sigma^{\prime(1,1)}_{\sigma\tau}+ \Sigma^{\prime\prime(1,1)}_{\sigma\tau} \right).
\end{split}
\end{equation}
In the one-loop, one-emission result the $\delta_{\sigma\tau\backslash
  n}$ indicates that the colour and anti-colour labels denoted by
$(c_n, \bar{c}_n)$ of the emitted gluon are at first merged and then
removed from the $\tau$-permutation such that a comparison to the
$\sigma$-permutation is possible.  \\ The $\rho$-coefficients again
denote colour flows which are suppressed in the number of colours. The
structure $\rho_\tau$ contains colour flows with a $1/N^2$
suppression, whereas $\rho_{\sigma\tau}$ stands for terms with only a
$1/N$ suppression and a colour connection in $\sigma$. For instance
consider
\begin{align}
\begin{gathered}
\vspace{-0.2cm}
\includegraphics[width=4.6cm]{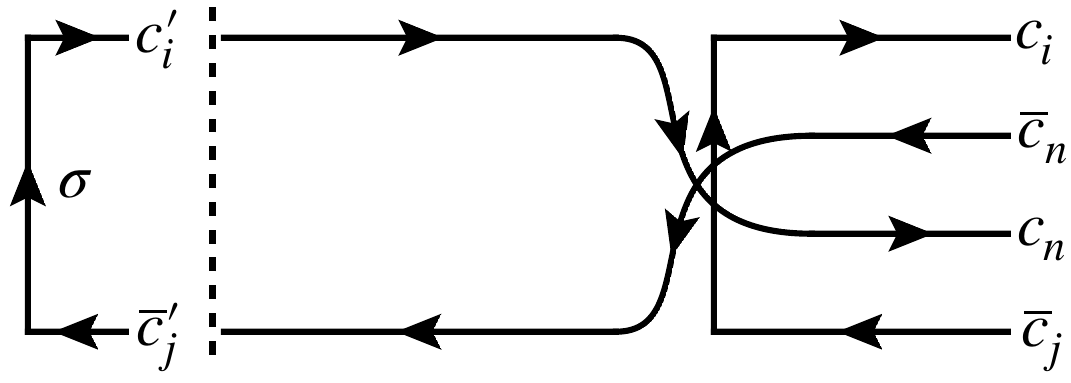}
\end{gathered}
\rightarrow \delta_{\sigma\tau\backslash n} \delta_{c_n\tau^{-1}(\overline{c}_n)} \delta_{c_i\sigma^{-1}(\overline{c}_j)} \ , &&
\end{align}
which is part of the matrix element $[\tau|\bold{T}_g \bold{T}_i
  \bold{T}_j|\sigma\rangle$ (cf. Eq. (\ref{A1.triple})), as an example
  of a contribution to $\rho_{\sigma\tau}$. All of the
  $\Sigma$-coefficients denote the various different structures which
  require swaps of colour labels to achieve matching permutations. The
  coefficient $\hat{\Sigma}_{\sigma\tau}^{(1,1)}$ includes single
  swaps with a colour connection in $\sigma$,
  $\tilde{\Sigma}_{\sigma\tau}^{(1,1)}$ represents colour flows with
  single swaps and a $1/N$ suppression due to U(1)-gluon exchange,
  while the suppression for the
  $\hat{\tilde{\Sigma}}_{\sigma\tau}^{(1,1)}$ contributions occurs due
  to U(1)-gluon emission. The latter can be exemplified by
\begin{align}
\begin{gathered}
\vspace{-0.2cm}
\includegraphics[width=4.6cm]{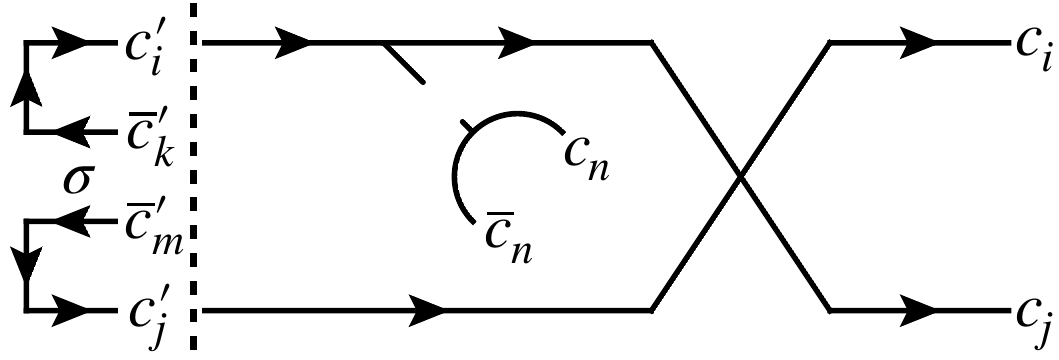}
\end{gathered}
\rightarrow \frac{1}{N} \delta_{c_n \tau^{-1}(\overline{c}_n)} \delta_{\sigma\tau_{(a,b)}\backslash n} \delta_{(a,b)(c_i,c_j)} \ . &&
\end{align}
This colour flow belongs to the matrix element $[\tau|(\bold{T}_i\cdot
  \bold{T}_j)\bold{T}_l |\sigma \rangle$ (cf. Eq. (\ref{A1.1e1l})).
  Just like at two-loop order, the coefficients $\Sigma^{\prime
    (1,1)}_{\sigma\tau}$ and $\Sigma^{\prime\prime
    (1,1)}_{\sigma\tau}$ denote double swaps of a single (anti-)
  colour label and swaps of four distinct (anti-)colour labels
  respectively. As an example for a contribution to
  $\Sigma^{\prime\prime (1,1)}_{\sigma\tau}$ consider
\begin{align}
\begin{gathered}
\vspace{-0.2cm}
\includegraphics[width=4.2cm]{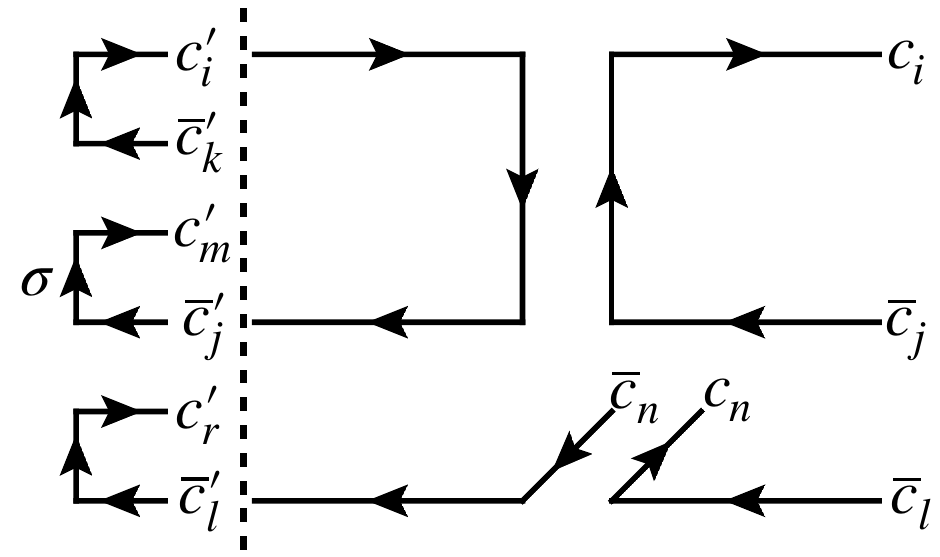}
\end{gathered}
\rightarrow \delta_{\sigma\tau_{(a,b)(c,d)}\backslash n} \delta_{(a,b)(\sigma^{-1}(\overline{c}_l),c_n)} \delta_{(c,d)(c_i \sigma^{-1}(\overline{c}_j))} \ , &&
\end{align}
which pertains to the matrix element $[\tau|(\bold{T}_i\cdot
  \bold{T}_j)\bold{T}_l |\sigma \rangle$ (cf. Eq. (\ref{A1.3leg})).
  The results of the colour structures can be used for the algorithm
  presented in \cite{DeAngelis:2020rvq} to generalise the sampling of
  emission flows including the one-loop corrections.

\section{Kinematic dependence and loop integrals}
\label{sec:Kernels}

As we aim at an evolution algorithm which is able to treat a large
class of observables differentially, we chose to relate the virtual
corrections systematically to phase space integrals, or integrals
which we can effectively treat as such. The advantage of this
procedure is that from these integrals we can single out an evolution
variable as appropriate to the observable we consider, and we can
devise local subtractions which make the cancellation of infrared
divergencies explicit. Exposing the remaining singularities leads to
an identification of the anomalous dimension by means of a counter
term required to obtain an overall finite cross section. In
particular, the anomalous dimension will be expressed in terms of
variables which can be linked to the real emission contributions. Such
a form is then very well suited for a Monte Carlo evolution algorithm
like the one outlined in \cite{DeAngelis:2020rvq}, and further
development thereof. The approach exploited here then also allows to
properly subtract collinear divergencies, and to make use of colour
conservation in order to factorise soft-collinear from (colour)
non-trivial soft, large-angle physics, see the discussion on collinear
subtractions and the ordering variable in \cite{Forshaw:2019ver}.

In the present work we solely consider soft gluon contributions and
use the Eikonal approximation throughout, though our approach will be
more generally applicable both to singular limits in the full QCD, as
well as to the Eikonal propagators encountered in the context of {\it
  e.g.} SCET \cite{Becher:2016mmh}. We also envisage that a
generalisation to the inclusion of collinear singularities will be
possible, however they are beyond the scope of the current work. In
the soft limit, for which we scale all of the real emission and
virtual gluon momenta by a common factor $\lambda$, we only keep the
leading singular term in an expansion around $\lambda\to 0$. We also
neglect the contribution of a soft $q\bar{q}$ pair, which in principle
needs to be taken into account and can easily be included using the
results we have presented here. We stress that this procedure
naturally guarantees that there will not be any approximation applied
in the case that soft gluons only couple via the three- and four-gluon
vertex, see also \cite{Angeles-Martinez:2015rna,
  Angeles-Martinez:2016dph}.

\subsection{Cutting rules}

For the case of the single gluon exchange, the relevant one-loop
integral can be performed using a contour integration, however in the
case of higher orders a more algorithmic treatment is desirable,
specifically if we aim for yet higher orders. To this extent we use
the Feynman tree theorem \cite{Feynman:1963ax,Catani:2008xa}, which we
extend to apply to the two-loop case, and to Eikonal propagators. Our
starting point is the relation between advanced and Feynman
propagators,
\begin{equation}
\frac{1}{\left[q^2-i0 (T\cdot q) |T\cdot q|\right]}=\frac{1}{[q^2+i0(T\cdot q)^2]}+2\pi i \delta(q^2) \theta(T\cdot q)\ ,
\label{eqs:cutidentity1}
\end{equation}
which is a modification of expressing the imaginary part of the Feynman propagator via
\begin{equation}
\frac{1}{[q^2-i0(T\cdot q)^2]}=\frac{1}{[q^2+i0(T\cdot q)^2]}+ 2\pi i \delta(q^2) \ .
\label{eqs:cutidentity2}
\end{equation}
It is noteworthy that, depending on the momentum routing of the
Feynman diagram, different cuts will appear when applying the Feynman
tree theorem which lead to different imaginary parts per cut
contribution, see \cite{Catani:2008xa} for instructive
examples. However, by combining the individual contributions the
results for the imaginary part will coincide regardless of the chosen
momentum routing. We have also highlighted the fact that the Feynman
$i0$ does need to carry mass dimensions, and that any projection onto
a time-like component of the loop momentum (one can choose $T^2 = 2$
for convenience\footnote{This choice guarantees that, in a frame where
$T=(\sqrt{2},\vec{0})$ we find that $1/(q^2+i 0 (T\cdot q)^2) = (1/(2
|\vec{q}|))(1/(q^0-|\vec{q}|+i 0 q^0)-1/(q^0+|\vec{q}|+i 0 q^0))$.})
is sufficient to guarantee the right deformation of the integration
contour around the poles in the complex $T\cdot k$ plane.

Eq.~(\ref{eqs:cutidentity1}) is at the heart of the Feynman tree
theorem, which uses the fact that an integral consisting of advanced
propagators only is bound to vanish as all propagator poles reside in
the upper half plane of the $T\cdot k$ plane. The full integral is
then expressed as a sum over cut diagrams containing any possible
configuration of (multiple) cut propagators. We will generalise this
method to the case of Eikonal propagators, for which we note that
\begin{equation}
  \frac{1}{2 p_i\cdot k - i 0 (T\cdot p_i)^2} = \frac{1}{2 p_i\cdot k + i 0 (T\cdot p_i)^2} + 2\pi i\ \delta(2 p_i\cdot k) 
  \label{eqs:cutidentity4}
\end{equation}
can be applied as cutting rule for an Eikonal propagator (with
$p_i>0$, which we should always assume here), for which the left hand
side does in fact admit a pole in the upper half plane. Cuts through
lines which have the soft gluon momentum running in the other
direction though, $1/(-2p_i\cdot k + i 0 (T\cdot p_i)^2)$ will not be
cut, since the propagator in the integral of interest has its pole
already in the upper half plane.

\begin{figure}
   \begin{center}
   \includegraphics[width=0.5\textwidth]{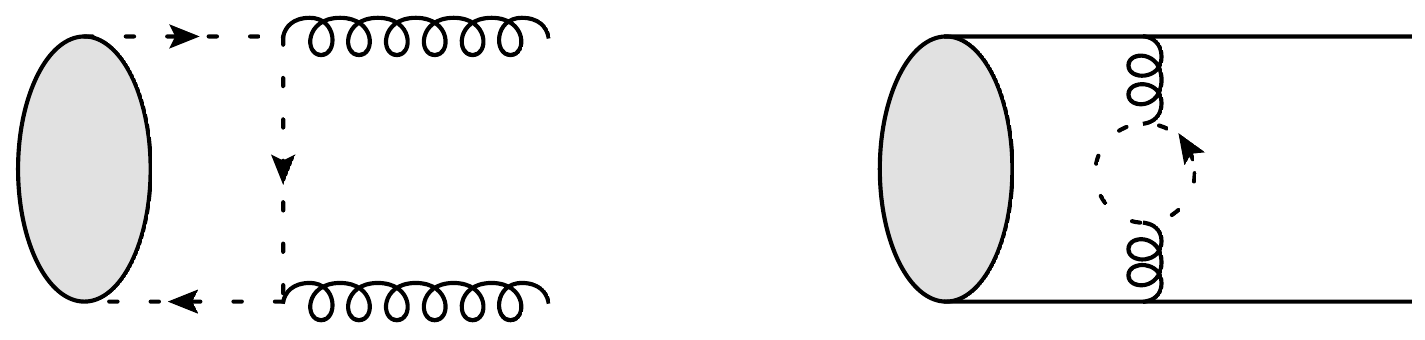}
   \end{center}
  \caption{\label{figs:noghosts}Examples of possible ghost
    contributions to the virtual corrections; due to the nature of the
    ghost coupling the diagram on the left hand side does not
    contribute in the soft limit. On the right hand side the ghost
    contribution for the gluon self energy is depicted, which needs to
    be included since it has the same momentum scaling as the three
    gluon vertex when all attached lines become soft.}
\end{figure}

Since we are focusing purely on soft effects, we will exemplify the
application of our method in a covariant gauge. In this case,
self-energy type diagrams are suppressed in the soft limit, and we
will only need to consider gluon exchange diagrams. Since we will
perform the analysis at the level of physical amplitudes, ghosts do
not appear as external lines, and diagrams with ghost exchanges with a
hard line are soft sub-leading owing to the momentum structure of the
ghost-gluon vertex. A notable exception is the gluon self energy, for
which ghost contributions need to be included, since those will
contribute in the double soft limit; on a similar note the three-gluon
vertex will not be approximated by taking the double soft limit, which
has already been noted in \cite{Angeles-Martinez:2016dph}. We depict
some diagrams with ghost contributions in Fig.~\ref{figs:noghosts}.

We do however point out that when one wants to include collinear
singular contributions a physical gauge is more advantageous. In this
case we further face poles stemming from the propagator term involving
the gauge vector $n$; in order to obtain the full $n$ (in)dependence
it is vital to also consider the linear propagator terms introduced in
the gluon propagator.  If we limit our consideration to the light-cone
gauge we can use the prescription \cite{Leibbrandt:Book}\footnote{We
  have slightly altered the form to be consistent with '0' meaning a
  dimensionless infinitesimal parameter, cf. the discussion on the
  Feynman propagator.}
\begin{equation}
  \frac{1}{n\cdot k}\to \frac{1}{n\cdot k + i 0\ n^* \cdot k} \ ,
\end{equation}
where $n^*=(n^0,-\vec{n})$ for $n=(n^0,\vec{n})$. This denominator
implementing the Mandelstam/Leibbrandt prescription has a pole in the
upper half-plane, if the spatial part of the loop momentum is
anti-parallel to $\vec{n}$, and in the lower half-plane otherwise. The
cutting rule is to assign a cut contribution $2\pi i \delta(n\cdot
k)\theta(\vec{n}\cdot \vec{k})$ to such a gauge denominator, and
accordingly we can build derivatives for higher powers.

\subsection{Application to Two-Loop Integrals}

\begin{figure}
 \begin{center}
   \includegraphics[width=0.9\textwidth]{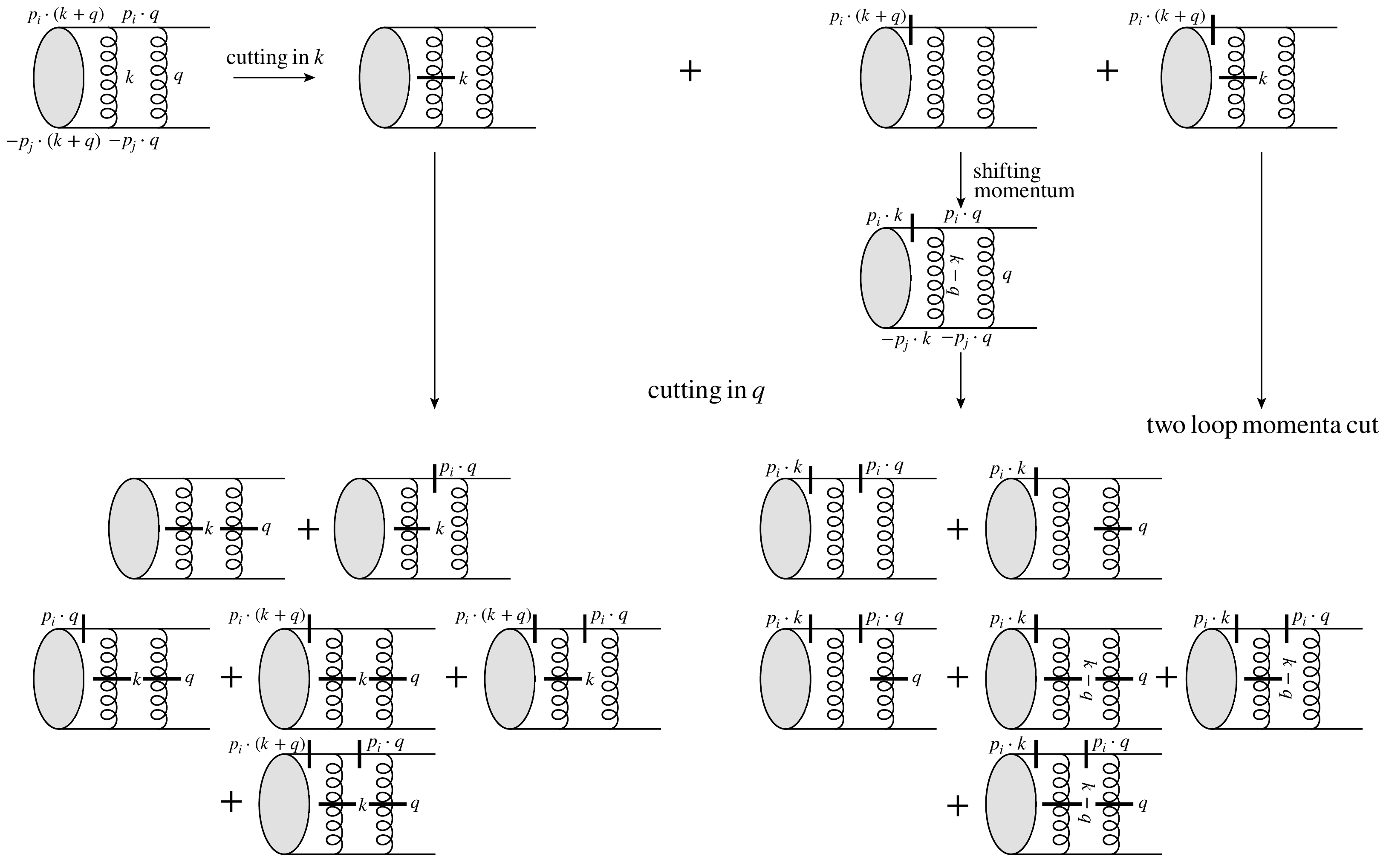}
   \end{center}
    \caption{\label{fig:2lworkflow}Illustration of the workflow when
      applying the Feynman tree theorem at the two-loop level. The
      short black lines symbolise cuts of the corresponding
      propagators.}
\end{figure}

The application of the Feynman tree theorem in the two-loop case is
not unique as compared to the one-loop case. However, we can apply it
to one of the one-loop sub-integrals, and then find a shift of the loop
momenta such that the second loop momentum after the transformation is
not affected by constraints of cuts from the previous application. If
this is not possible, we terminate the algorithm. In this way we
maximise the number of contributions of the two-loop integral which
can directly be cast into a form of a double-emission phase space-type
integral. After having cut in a first loop integration $k$ we
determine, for each term in the sum over cuts, if we can shift the
loop momenta $k,q\to k',q'$ in such a way that one loop integration,
say $q'$, only involves propagators and that no $\delta$ or
$\theta$-function involves $q'$. Then the $q'$ integration can again
be treated with the Feynman tree theorem. If this is not possible to
perform such a momentum shift and both loop momenta are constrained by
the $\delta$-functions already after the first application, or if the
transformation $k,q\to k',q'$ would enforce a negative-energy
$\theta$-function which has originated from the previous cut, the
procedure terminates. The algorithm we outline here has been automated
and we have used it to process all integrals required to calculate the
two-loop soft integrals. We also exemplify the procedure in
Fig.~\ref{fig:2lworkflow}, which shows a detailed workflow for one of
the topologies we have considered.
Starting from the results for the two-loop integrals we then can
choose an explicit parametrisation of the loop momentum in order to
make contact with the phase space integrals over the real emission,
and the resolution scale of the observable. At this stage, however, we
only quote the (non-vanishing) cut contributions, including those
which give rise to imaginary parts and will address a more concrete
parametrisation tailored for specific observables in a future
work. All of the relevant integrals in the two-loop, soft gluon case,
are listed in Appendix~\ref{sec:Integrals}. The loop integrals which 
correspond to the $\Omega^{(2)}$-coefficients involving two hard 
lines are given by
\begin{align}
&\Omega^{(2)}_{ij}= -(p_i\cdot p_j)(p_i\cdot p_j) \, I^{(ij)}_1 \ , \nonumber \\
&\tilde{\Omega}^{(2)}_{ij}=-(p_i\cdot p_j)(p_i\cdot p_j) \, I^{(ij)}_2 \ , \nonumber \\
&\Omega^{(2)}_{ij, \mathrm{vertex-corr.}}=C_A \, (p_i\cdot p_j)\left[2 I^{(ij)}_3- I^{(ij)}_4\right] \ ,  \\ \nonumber
&\Omega^{(2)}_{ij, \mathrm{self-en.|gluon+ghost}}= 2 C_A\left[(4-2d)\, I^{(ij)}_5 -2(p_i\cdot p_j)\, I^{(ij)}_3- (p_i\cdot p_j) \, I^{(ij)}_6\right] \\ \nonumber
&\Omega^{(2)}_{ij, \mathrm{self-en.|fermion}}=4n_f \left[2 I^{(ij)}_5-\frac{1}{2} (p_i\cdot p_j) I^{(ij)}_3 -(p_i\cdot p_j) I^{(ij)}_6\right] \ ,
\end{align}
and involving three hard lines we find
\begin{align}
&\Omega^{(2)}_{ijl}=-(p_i\cdot p_l)(p_i\cdot p_j) \, I^{(ijl)}_1\ , \\ \nonumber
&\hat{\Omega}^{(2)}_{ijl}=2 \left[(p_j\cdot p_l)\, I^{(ijl)}_2 - (p_i\cdot p_j) \, I^{(ijl)}_3 -(p_i\cdot p_l) \, I^{(ijl)}_4 \right] \ .
\end{align}

\subsection{Application to Higher Propagator Powers}

In the two-loop case we will also need to consider propagators
at higher powers, in particular when a self-energy correction is
inserted on an exchanged gluon.  In order to find an analogous rule
for treating propagators of higher powers, we consider the derivative
of equation (\ref{eqs:cutidentity2}) with respect to the momentum
$q^\mu$
\begin{equation}
\frac{q^\mu-i0T^\mu(T\cdot q)}{[q^2-i0(T\cdot q)^2]^2}-\frac{q^\mu+i0 T^\mu (T \cdot q)}{[q^2+i0 (T\cdot q)^2]^2}=- i \pi \frac{\partial}{\partial q^\mu} \delta(q^2) \ .
\end{equation}
The derivative of the $\delta$-function with respect to a momentum
component can be left as a formal object until we can simplify this
later by means of integration by parts. An alternative which we
consider is to eliminate the $T$-dependent term in the numerator,
contract with some vector $S^\mu$, which has the property $S\cdot
T=0$, i.e. $S^\mu$ is space-like ($S^2<0$) to obtain
\begin{equation}
\begin{split}
\frac{1}{[q^2-i0 (T\cdot q)^2]^2}-\frac{1}{[q^2+i0(T\cdot q)^2]^2}&=-i\pi \frac{S_\mu}{S\cdot q} \frac{\partial}{\partial q^\mu} \delta(q^2) \\
&=-2i\pi \delta'(q^2)\ ,
\end{split}
\end{equation}
where we define $\delta'(q^2)\equiv \frac{\partial}{\partial
  q^2}\delta(q^2)$. This identity can then be supplemented by placing
the double pole again in the upper half plane and we effectively use
\begin{equation}
  \frac{1}{[q^2-i0(T\cdot q)|T\cdot q|]^2}-\frac{1}{[q^2+i0(T\cdot q)^2]^2}= - 2i\pi \theta(T\cdot q) \delta'(q^2) \ .
    \label{eqs:cutidentity3}
\end{equation}
Clearly this procedure can continue to even higher powers of
propagators. In our case we require to apply Eq.~(\ref{eqs:cutidentity3})
when considering a self-energy insertion on an exchanged gluon
line. We note that, if we have not expressed the self energy in its
integrated form but do want to apply our cutting procedure, then we
will face $\delta'$-distributions possibly in both loop
integrations. We can ultimately only resolve these constraints using
integration by parts once we have expressed the loop momenta in a
certain parametrisation. An example of this procedure is to write
\begin{equation}
q^\mu=q_+\overline{n}^\mu+ \frac{q^2-q_\perp^2}{2q_+ (n\cdot \overline{n})}n^\mu+q_\perp^\mu \ , 
\label{loopmomparametrisation}
\end{equation}
as this leads to the following expression for the derivative with respect to $q^2$
\begin{equation}
\frac{\partial q^\mu}{\partial q^2}= \frac{n^\mu}{2q_+(n\cdot \overline{n})}=\frac{n^\mu}{2(n\cdot q)}\ ,
\end{equation}
with $q_+=\frac{n\cdot q}{n\cdot\overline{n}}$. No reference then
needs to be made to the additional vectors $T^\mu$ and $S^\mu$ which
we have been resorting to in setting up the derivative rule. In
App.~\ref{sec:SelfEnergy} we give expressions for the relevant
integrals using this procedure.

\subsection{One-loop, one-emission contributions}
The integrands of the one-loop, one-emission diagrams have been treated
with the modified Feynman tree theorem for Eikonal propagators. To
ensure the applicability of the FTT we checked for each of the
diagrams that the degrees of the polynomial in the loop-momentum of
the propagators is bigger or equal to the polynomial of the numerator
structure plus two. For the diagram involving the triple gluon vertex
a tensor reduction can be performed prior to the application of the
FTT. The cut-contributions which are obtained through this
procedure for the one-loop, one-emission diagrams are listed in
Appendix \ref{app:FTT11}. The coefficients $\Omega^{(1,1)}$ for two
hard lines are given by
\begin{align}
&\Omega^{(1,1)}_{ij}= (-i) \frac{(p_i\cdot p_j)}{p_i\cdot q} (p_i\cdot \varepsilon^*(q))\,  I^{(ij)}_7 \ , \nonumber\\
&\tilde{\Omega}^{(1,1)}_{ij}=(-i) (p_i\cdot p_j)(p_i\cdot \varepsilon^*(q))\, I^{(ij)}_8 \ ,\\
&\overline{\Omega}^{(1,1)}_{ij}= i \Bigg\{(p_j\cdot q) (p_i\cdot \varepsilon^*(q))\, I^{(ij)}_9- \frac{p_i\cdot q}{p_j\cdot q}(p_j\cdot \varepsilon^*(q)) \, I^{(ij)}_{10}\nonumber\\ \nonumber
&\hspace{1cm}- \frac{p_i\cdot p_j}{2} \left[\frac{(p_i\cdot \varepsilon^*(q))}{p_i\cdot q}+ \frac{(p_j\cdot \varepsilon^*(q))}{p_j\cdot q}\right] I^{(ij)}_7\Bigg\} \ , \\
&\Omega^{(1,1)}_{ij, \mathrm{vertex-corr.}}=C_A (p_i\cdot \varepsilon^*(q)) \left[I^{(ij)}_{10}- \frac{2}{p_i\cdot q} I_0^{(ij)}\right]\ , \nonumber
\end{align}
where the loop integral described by $I^{(ij)}_0$ is scale-less and 
if the FTT is used one finds that it vanishes exactly for $p_i^0>0$ 
and on-shell external momenta. For the one-loop/one-emission contribution 
with three hard lines one gets the one-loop integral $\Omega^{(1)}_{il}$ (cf. Eq. (\ref{eqs:gammadefinition}))
\begin{align}
\Omega^{(1,1)}_{ijl}=\frac{p_i\cdot p_l}{p_j\cdot q} (p_j\cdot \varepsilon^*(q)) \, \Omega^{(1)}_{il} \ .
\end{align}
\section{Conclusion and Outlook}
\label{sec:Outlook}

In this paper we have been providing the basic ingredients to
formulate soft gluon evolution at the next-to-leading order, using the
colour flow basis, as well as the Feynman tree theorem, to express the
relevant one-loop, one-emission and two-loop diagrams.

This will allow to perform the resummation of non-global observables
at the next-to-leading logarithmic order, and provides crucial insight
to the design of amplitude evolution algorithms and parton showers
beyond the currently adopted approximations. In particular we have
taken an entirely new approach towards virtual corrections, being able
to cast them into a form in which we can make the cancellation of
infrared divergencies explicit and determine the imaginary parts of
the loop integrals in a way that we control the kinematic regions from
which they originate. Results on this, assuming certain orderings,
will be published elsewhere, while we have here presented the virtual
corrections in a generic manner using a new method of applying the FTT
to the two-loop case and to Eikonal propagators. All of our results
are carried out in dimensional regularisation such that upon
performing the actual integrals one could recover the poles in an
expansion in $\epsilon\to 0$.

The analysis of the colour structure allows us to adopt a perturbative
treatment of the amplitude evolution operators, or to explore more
sophisticated Monte Carlo methods in colour space. Our analysis
definitely highlights that a dipole-type picture is not anymore
sufficient provided one concentrates not only on the strict leading
colour approximation, but requires to include all colour diagonal
contributions, an approximation which has turned out to be crucial
within the leading-order context already, and certainly if one
attempts to include collinear physics which probes such effects
effectively as the difference between $C_A/2$ and $C_F$.

\section*{Acknowledgments}

We are grateful to Thomas Becher, Jeff Forshaw and Maximillian
L\"oschner for fruitful discussions. We would also like to thank Jeff
Forshaw for detailed and valuable comments on the present manuscript.
We are grateful for the kind hospitality of Mainz Institute for
Theoretical Physics (MITP) of the DFG Cluster of Excellence PRISMA$^+$
(project ID 39083149), where some of this work has been carried out,
as well as the Erwin Schr\"odinger Institute at Vienna for support
through a Research in Teams program.

This work has also been supported in part by the European
Union’s Horizon 2020 research and innovation programme as part of the
Marie Skłodowska-Curie Innovative Training Network MCnetITN3 (grant
agreement no. 722104). SP acknowledges partial support by the COST
actions CA16201 ``PARTICLEFACE'' and CA16108 ``VBSCAN''. \\
Figures have been prepared using JaxoDraw \cite{Binosi:2008ig}.

\newpage

\appendix

\section{Details on colour structures}
\label{sec:ColourCorrelators}
In this appendix we quote the main results on the decomposition of the
colour structures in the one-loop, one-emission as well as two loop
cases, translating the colour structures into the colour flow basis
and quoting the final result in terms of swaps and colour flows added
after the emission. We quote those per diagram, as we have extracted
them to separate the colour structures from the kinematic dependence.
\subsection{One-loop, one-emission colour structures}
\begin{itemize}
\item{Gluon emission after the exchange:}
\end{itemize}

\small
\begin{equation}
\begin{split}
[\tau|\bold{T}_i (\bold{T}_i \cdot \bold{T}_j) |\sigma\rangle&= N^2 \delta_{\sigma \tau \backslash n} \\
&\left[ \frac{1}{N^2} (\lambda_i- \bar{\lambda}_i) \left(\lambda_i \bar{\lambda}_j \delta_{c_i \sigma^{-1}(\overline{c}_j)} \delta_{c_n \tau^{-1}(\overline{c}_n)}+\bar{\lambda}_i \lambda_j \delta_{c_j \sigma^{-1}(\overline{c}_i)} \delta_{c_n \tau^{-1}(\overline{c}_n)}\right) \right. \\
& \left. + \frac{1}{N^4} (\lambda_i- \bar{\lambda}_i)^2 (\lambda_j-\bar{\lambda}_j)  \delta_{c_n \tau^{-1}(\overline{c}_n)} \right] \\
&+ \sum_{(a,b)} \delta_{\sigma \tau_{(a,b)} \backslash n} \\
&\bigg[N \left(\bar{\lambda}_i^2 \lambda_j \delta_{c_j \sigma^{-1}(\overline{c}_i)} \delta_{(a,b)(c_j, c_n)} - \lambda_i^2 \bar{\lambda}_j \delta_{c_i \sigma^{-1}(\overline{c}_j)}  \delta_{(a,b)(c_i, c_n)} \right.  \\
& \left. + \lambda_i \bar{\lambda}_i \lambda_j \delta_{c_i \sigma^{-1}(\overline{c}_j)} \delta_{(a,b)(\sigma^{-1}(\overline{c}_i), c_n)} - \lambda_i \bar{\lambda}_i \lambda_j \delta_{c_j \sigma^{-1}(\overline{c}_i)} \delta_{(a,b)(c_i, c_n)}\right)  \\ 
& + \frac{1}{N} \delta_{c_n \tau^{-1}(\overline{c}_n)} (\lambda_i- \bar{\lambda}_i) \left( \lambda_i \bar{\lambda}_j \delta_{(a,b)(c_i, \sigma^{-1}(\overline{c}_j))} + \bar{\lambda}_i \lambda_j  \delta_{(a,b)(c_j, \sigma^{-1}(\overline{c}_i))} \right. \\
& \left. - \lambda_i \lambda_j  \delta_{(a,b)(c_i, c_j)} - \bar{\lambda}_i \bar{\lambda}_j \delta_{(a,b)(\sigma^{-1}(\overline{c}_j), \sigma^{-1}(\overline{c}_i))} \right)  \\
& -\frac{1}{N} (\lambda_i- \bar{\lambda}_i) (\lambda_j- \bar{\lambda}_j)\left(\lambda_i  \delta_{(a,b)(c_i, c_n)} - \bar{\lambda}_i  \delta_{(a,b) (\sigma^{-1}(\overline{c}_i), c_n)}\right)\bigg] \\
& + \sum_{(a,b)} \sum_{(b,c)} \delta_{\sigma \tau_{(a,b)(b,c)}\backslash n} \\
&\left[\lambda_i^2  \lambda_j \delta_{(a,b)(c_n, c_i)} \delta_{(b,c)(c_i, c_j)} + \bar{\lambda}_i^2 \lambda_j \delta_{(a,b)(c_n, c_j)} \delta_{(b,c)(c_j, \sigma^{-1}(\overline{c}_i))}  \right. \\
& \left. - \lambda_i^2  \bar{\lambda}_j \delta_{(a,b)(c_n, c_i)} \delta_{(b,c)(c_i, \sigma^{-1}(\overline{c}_j))} - \bar{\lambda}_i^2 \bar{\lambda}_j \delta_{(a,b)(c_n, \sigma^{-1}(\overline{c}_j))} \delta_{(b,c)(\sigma^{-1}(\overline{c}_j), \sigma^{-1}(\overline{c}_i))} \right] \\
& + \sum_{(a,b)} \sum_{(c,d)}  \delta_{\sigma \tau_{(a,b)(c,d)}\backslash n}  \lambda_i \bar{\lambda}_i \\
&\left[\delta_{(a,b)(\sigma^{-1}(\overline{c}_i), c_n)} \left(\bar{\lambda}_j \delta_{(c,d)(c_i, \sigma^{-1}(\overline{c}_j))} - \lambda_j \delta_{(c,d)(c_i, c_j)}\right) \right. \\
& \left.+ \delta_{(a,b)(c_i, c_n)} \left(\bar{\lambda}_j \delta_{(c,d)(\sigma^{-1}(\overline{c}_i), \sigma^{-1}(\overline{c}_j))} - \lambda_j \delta_{(c,d)(c_j, \sigma^{-1}(\overline{c}_i))} \right) \right] 
\end{split}
\label{A1.1l1e}
\end{equation}
\normalsize
\newpage
\begin{itemize}
\item{Triple gluon vertex:}
\end{itemize}

\small
\begin{equation}
\begin{split}
[\tau| \bold{T}_g \bold{T}_i \bold{T}_j |\sigma\rangle&= \sqrt{T_R} \left\{ \delta_{\sigma \tau \backslash n} \delta_{c_n \tau^{-1}(\overline{c}_n)} \left[-\lambda_i \bar{\lambda}_j \delta_{c_i \sigma^{-1}(\overline{c}_j)}+ \bar{\lambda}_i \lambda_j \delta_{c_j \sigma^{-1}(\overline{c}_i)} \right] \right.\\
&\left. + N \sum_{(a,b)} \delta_{\sigma \tau_{(a,b)}\backslash n} \left[ \lambda_i \bar{\lambda}_j \delta_{c_i \sigma^{-1}(\overline{c}_j)} \delta_{(a,b)(c_i, c_n)} - \bar{\lambda}_i \lambda_j \delta_{c_j \sigma^{-1}(\overline{c}_i)} \delta_{(a,b)(c_j, c_n)} \right] \right. \\
&\left. + \sum_{(a,b)} \sum_{(b,c)} \delta_{\sigma \tau_{(a,b)(b,c)} \backslash n} \right. \\
&\left. \left[\lambda_i \bar{\lambda}_j \left(\delta_{(a,b)(c_n, c_i)} \delta_{(b,c)(c_i, \sigma^{-1}(\overline{c}_j))} - \delta_{(a,b)(c_n, \sigma^{-1}(\overline{c}_j))} \delta_{(b,c)(\sigma^{-1}(\overline{c}_j), c_i)}\right)  \right. \right. \\
& \left. \left. + \bar{\lambda}_i \lambda_j \left( \delta_{(a,b)(c_n, \sigma^{-1}(\overline{c}_i))} \delta_{(b,c)(\sigma^{-1}(\overline{c}_i), c_j)} - \delta_{(a,b)(c_n, c_j)} \delta_{(b,c)(c_j, \sigma^{-1}(\overline{c}_i))} \right) \right. \right. \\
&\left. \left. +\lambda_i \lambda_j \left( \delta_{(a,b)(c_n, c_j)} \delta_{(b,c)(c_j, c_i)}  - \delta_{(a,b)(c_n, c_i)} \delta_{(b,c)(c_i, c_j)} \right) \right. \right. \\
& \left. \left. + \bar{\lambda}_i \bar{\lambda}_j \left(\delta_{(a,b)(c_n, \sigma^{-1}(\overline{c}_j))} \delta_{(b,c)(\sigma^{-1}(\overline{c}_j), \sigma^{-1}(\overline{c}_i))}\right. \right. \right.  \\
&\left. \left. \left.  - \delta_{(a,b)(c_n, \sigma^{-1}(\overline{c}_i))}  \delta_{(b,c)(\sigma^{-1}(\overline{c}_i), \sigma^{-1}(\overline{c}_j))} \right) \right] 
\right\}  
\label{A1.triple}
\end{split}
\end{equation}
\normalsize
\vspace{1cm}
\begin{itemize}
\item{Gluon emission and exchange involving three hard lines:}
\end{itemize}

\small
\begin{equation}
\begin{split}
[\tau|(\bold{T}_i\cdot \bold{T}_j)\bold{T}_l |\sigma \rangle=&N^2 \delta_{\sigma \tau \backslash n} \delta_{c_n\tau^{-1}(\overline{c}_n)} \left[ \frac{1}{N^4} (\lambda_i - \bar{\lambda}_i) (\lambda_j- \bar{\lambda}_j) (\lambda_l- \bar{\lambda}_l) \right. \\
&\left. + \frac{1}{N^2} \left( \bar{\lambda}_i \lambda_j (\lambda_l - \bar{\lambda}_l) \delta_{c_j\sigma^{-1}(\overline{c}_i)} + \lambda_i \bar{\lambda}_j (\lambda_l - \bar{\lambda}_l) \delta_{c_i \sigma^{-1}(\overline{c}_j)} \right) \right] \\
&+ \sum_{(a,b)} \delta_{\sigma \tau_{(a,b)} \backslash n}  \left[ N \delta_{(a,b)(\sigma^{-1}(\overline{c}_l), c_n)} \left(\lambda_i \bar{\lambda}_j  \bar{\lambda}_l  \delta_{c_i \sigma^{-1}(\overline{c}_j)}  + \bar{\lambda}_i \lambda_j \bar{\lambda}_l \delta_{c_j \sigma^{-1}(\overline{c}_i}) \right) \right. \\
&\left. - N  \delta_{(a,b)(c_l, c_n)} \left(\lambda_i \bar{\lambda}_j \lambda_l \delta_{c_i \sigma^{-1}(\overline{c}_j)} +\bar{\lambda}_i \lambda_j \lambda_l \delta_{c_j \sigma^{-1}(\overline{c}_i)} \right) \right. \\
&\left. +\frac{1}{N} (\lambda_i - \bar{\lambda}_i) (\lambda_j - \bar{\lambda}_j) \left(\bar{\lambda}_l \delta_{(a,b)(\sigma^{-1}(\overline{c}_l), c_n)} -  \lambda_l  \delta_{(a,b)(c_l, c_n)}\right) \right.\\
&\left. +\frac{1}{N} \delta_{c_n \tau^{-1}(\overline{c}_n)}(\lambda_l - \bar{\lambda}_l) \left(\lambda_i \bar{\lambda}_j  \delta_{(a,b)(c_i, \sigma^{-1}(\overline{c}_j))} + \bar{\lambda}_i \lambda_j  \delta_{(a,b)(\sigma^{-1}(\overline{c}_i), c_j)} \right. \right. \\
&\left. \left. - \lambda_i \lambda_j \delta_{(a,b)(c_i, c_j)}  - \bar{\lambda}_i \bar{\lambda}_j  \delta_{(a,b)(\sigma^{-1}(\overline{c}_i), \sigma^{-1}(\overline{c}_j))} \right) \right] \\
&+ \sum_{(a,b)} \sum_{(c,d)} \delta_{\sigma \tau_{(a,b)(c,d)}\backslash n} \\
&\left[- \lambda_i \bar{\lambda}_l  \delta_{(a,b)(\sigma^{-1}(\overline{c}_l), c_n)} \left(\lambda_j \delta_{(c,d)(c_i, c_j)} - \bar{\lambda}_j \delta_{(c,d)(c_i, \sigma^{-1}(\overline{c}_j))} \right) \right. \\
&\left. + \bar{\lambda}_i \bar{\lambda}_l \delta_{(a,b)(\sigma^{-1}(\overline{c}_l), c_n)} \left(\lambda_j \delta_{(c,d)(\sigma^{-1}(\overline{c}_i), c_j))} -\bar{\lambda}_j \delta_{(c,d)(\sigma^{-1}(\overline{c}_i), \sigma^{-1}(\overline{c}_j))} \right) \right. \\
&\left. +  \lambda_i  \lambda_l \delta_{(a,b)(c_l, c_n)} \left(\lambda_j \delta_{(c,d)(c_i, c_j)} - \bar{\lambda}_j \delta_{(c,d)(c_i, \sigma^{-1}(\overline{c}_j))}\right) \right. \\
&\left. - \bar{\lambda}_i  \lambda_l \delta_{(a,b)(c_l, c_n)} \left(\lambda_j \delta_{(c,d)(\sigma^{-1}(\overline{c}_i), c_j)} - \bar{\lambda}_j \delta_{(c,d)(\sigma^{-1}(\overline{c}_i), \sigma^{-1}(\overline{c}_j))} \right) \right] 
\end{split}
\label{A1.3leg}
\end{equation}
\normalsize
\newpage
\begin{itemize}
\item{Gluon emission prior to exchange:}
\end{itemize}

\small
\begin{equation}
\begin{split}
[\tau|(\bold{T}_i \cdot \bold{T}_j) \bold{T}_i |\sigma\rangle&= N^2 \delta_{\sigma \tau \backslash n} \delta_{c_n \tau^{-1}(\overline{c}_n)} \left[\frac{1}{N^2} \left(\lambda_i \bar{\lambda}_i \lambda_j \delta_{c_j \sigma^{-1}(\overline{c}_i)} -\lambda_i \bar{\lambda}_i \bar{\lambda}_j \delta_{c_i \sigma^{-1}(\overline{c}_j)}\right) \right. \\
& \left. +\frac{1}{N^4} (\lambda_i-\bar{\lambda}_i)^2 (\lambda_j - \bar{\lambda}_j)  \right] \\
&+\sum_{(a,b)} \delta_{\sigma \tau_{(a,b)}\backslash n} \\
&\left[N \lambda_i \bar{\lambda}_i \left(\bar{\lambda}_j \delta_{c_i \sigma^{-1}(\overline{c}_j)} \delta_{(a,b)(\sigma^{-1}(\overline{c}_i), c_n)} - \lambda_j  \delta_{c_j \sigma^{-1}(\overline{c}_i)} \delta_{(a,b)(c_i, c_n)}\right) \right. \\
&\left. +\frac{1}{N} \delta_{c_n \tau^{-1}(\overline{c}_n)} (\lambda_i- \bar{\lambda}_i)  \left(\bar{\lambda}_j \lambda_i  \delta_{(a,b)(c_i, \sigma^{-1}(\overline{c}_j))} + \bar{\lambda}_i \lambda_j  \delta_{(a,b)(\sigma^{-1}(\overline{c}_i), c_j)} \right. \right. \\
&\left. \left. - \lambda_i \lambda_j \delta_{(a,b)(c_i, c_j)} - \bar{\lambda}_i \bar{\lambda}_j \delta_{(a,b)(\sigma^{-1}(\overline{c}_i), \sigma^{-1}(\overline{c}_j))} \right) \right. \\
&\left. - \frac{1}{N} \lambda_i (\lambda_i- \bar{\lambda}_i) (\lambda_j- \bar{\lambda}_j) \delta_{(a,b)(c_i, c_n)}\right. \\
&\left.  + \frac{1}{N} \bar{\lambda}_i (\lambda_i- \bar{\lambda}_i) (\lambda_j- \bar{\lambda}_j) \delta_{(a,b)(\sigma^{-1}(\overline{c}_i), c_n)} \right] \\
&+\sum_{(a,b)} \sum_{(b,c)} \delta_{\sigma \tau_{(a,b)(b,c)} \backslash n} \\
&\left[\lambda_i^2  \lambda_j \delta_{(a,b)(c_n, c_j)}  \delta_{(b,c)(c_j, c_i)}  + \bar{\lambda}_i^2 \lambda_j \delta_{(a,b)(c_n, \sigma^{-1}(\overline{c}_i))} \delta_{(b,c)(\sigma^{-1}(\overline{c}_i), c_j)} \right. \\
&\left. - \lambda_i^2 \bar{\lambda}_j  \delta_{(a,b)(c_n, \sigma^{-1}(\overline{c}_j))} \delta_{(b,c)(\sigma^{-1}(\overline{c}_j), c_i)}\right.  \\
&\left. - \bar{\lambda}_i^2 \bar{\lambda}_j  \delta_{(a,b) (c_n, \sigma^{-1}(\overline{c}_i))} \delta_{(b,c)(\sigma^{-1}(\overline{c}_i), \sigma^{-1}(\overline{c}_j))}  \right. \\
&\left. - \lambda_i \bar{\lambda}_i \lambda_j \delta_{c_j \sigma^{-1}(\overline{c}_i)} \delta_{(a,b)(c_n, c_i)} \delta_{(b,c)(c_i, c_j)} \right. \\
&\left. + \lambda_i \bar{\lambda}_i \bar{\lambda}_j \delta_{c_i \sigma^{-1}(\overline{c}_j)} \delta_{(a,b)(c_n, c_i)} \delta_{(b,c)(c_i, \sigma^{-1}(\overline{c}_i))}\right] \\
&+\sum_{(a,b)} \sum_{(c,d)} \delta_{\sigma \tau_{(a,b)(c,d)} \backslash n} \lambda_i \bar{\lambda}_i \\
&\left[ \bar{\lambda}_j \left( \delta_{(a,b)(\sigma^{-1}(\overline{c}_i), c_n)} \delta_{(c,d)(c_i, \sigma^{-1}(\overline{c}_j))}- \delta_{(a,b)(c_i, c_n)}  \delta_{(c,d)(\sigma^{-1}(\overline{c}_i), \sigma^{-1}(\overline{c}_j))} \right) \right. \\
&\left. - \lambda_j  \left( \delta_{(a,b)(\sigma^{-1}(\overline{c}_i), c_n)} \delta_{(c,d)(c_i, c_j)} + \delta_{(a,b)(c_i, c_n)} \delta_{(c,d)(\sigma^{-1}(\overline{c}_i), c_j)} \right) \right] 
\end{split}
\label{A1.1e1l}
\end{equation}
\normalsize
%%%%%%%%%%%%%%%%%%%
\newpage
\subsection{Two-loop structures}
\begin{itemize}
\item{Double gluon exchange:}
\end{itemize}

\small
\begin{equation}
\begin{split}
[\tau|(\bold{T}_i \cdot \bold{T}_j)(\bold{T}_i \cdot \bold{T}_j)|\sigma \rangle &= N^2 \delta_{\sigma \tau}\\
& \bigg[ \lambda_i^2 \bar{\lambda}_j^2 \delta_{c_i \sigma^{-1}(\overline{c}_j)} + \bar{\lambda}_i^2 \lambda_j^2 \delta_{c_j \sigma^{-1}(\overline{c}_i)} + 2 \lambda_i \bar{\lambda}_i \lambda_j \bar{\lambda}_j  \delta_{c_i \sigma^{-1}(\overline{c}_j)} \delta_{c_j \sigma^{-1}(\overline{c}_i)}  \\
& + \frac{2}{N^2} (\lambda_i - \bar{\lambda}_i) (\lambda_j - \bar{\lambda}_j) \left(\bar{\lambda}_i \lambda_j \delta_{c_j \sigma^{-1}(\overline{c}_i)} + \lambda_i \bar{\lambda}_j \delta_{c_i \sigma^{-1}(\overline{c}_j)} \right) \\
& + \frac{1}{N^2} \left(\lambda_i^2 \lambda_j^2 + \bar{\lambda}_i^2 \bar{\lambda}_j^2\right) + \frac{1}{N^4} (\lambda_i - \bar{\lambda}_i)^2 (\lambda_j- \bar{\lambda}_j)^2 \bigg] \\
& + N \sum_{(a,b)} \delta_{\sigma \tau_{(a,b)}} \left[\lambda_i^2 \bar{\lambda}_j^2 \delta_{(a,b)(c_i, \sigma^{-1}(\overline{c}_j))} + \bar{\lambda}_i^2 \lambda_j^2 \delta_{(a,b)(\sigma^{-1}(\overline{c}_i), c_j)}  \right. \\
& \left. + \frac{2}{N^2} (\lambda_i - \bar{\lambda}_i) (\lambda_j - \bar{\lambda}_j) \left(\lambda_i \bar{\lambda}_j \delta_{(a,b)(c_i, \sigma^{-1}(\overline{c}_j))} + \bar{\lambda}_i \lambda_j \delta_{(a,b)(\sigma^{-1}(\overline{c}_i), c_j)} \right. \right. \\
& \left. \left.  - \lambda_i \lambda_j \delta_{(a,b)(c_i, c_j)} - \bar{\lambda}_i \bar{\lambda}_j \delta_{(a,b)(\sigma^{-1}(\overline{c}_i), \sigma^{-1}(\overline{c}_j))} \right) \right. \\
& \left. - \lambda_i \bar{\lambda}_i \left(\lambda_j^2 \delta_{(a,b)(c_i, c_j)} \delta_{c_j \sigma^{-1}(\overline{c}_i)} + \bar{\lambda}_j^2 \delta_{(a,b)(c_i, \sigma^{-1}(\overline{c}_i))} \delta_{c_i \sigma^{-1}(\overline{c}_j)} \right) \right. \\
& \left. - \lambda_j \bar{\lambda}_j \left(\lambda_i^2 \delta_{(a,b)(c_i, c_j)} \delta_{c_i \sigma^{-1}(\overline{c}_j)} + \bar{\lambda}_i^2 \delta_{(a,b)(c_j, \sigma^{-1}(\overline{c}_j))} \delta_{c_j \sigma^{-1}(\overline{c}_i)} \right) \right. \\
& \left. + 2 \lambda_i \bar{\lambda}_i \lambda_j \bar{\lambda}_j \left(\delta_{(a,b)( \sigma^{-1}(\overline{c}_i), c_j)} \delta_{c_i \sigma^{-1}(\overline{c}_j)} + \delta_{(a,b)(c_i, \sigma^{-1}(\overline{c}_j))} \delta_{c_j \sigma^{-1}(\overline{c}_i)} \right) \right] \\
& -\sum_{(a,b)} \sum_{(b,c)} \delta_{\sigma \tau_{(a,b)(b,c)}} \\
&\left[ \lambda_i \bar{\lambda}_i \lambda_j^2 \left(\delta_{(a,b)(c_i, \sigma^{-1}(\overline{c}_i))}  \delta_{(b,c)(\sigma^{-1}(\overline{c}_i), c_j)}  + \delta_{(a,b)(c_i, c_j)} \delta_{(b,c)(c_j, \sigma^{-1}(\overline{c}_i))} \right) \right. \\
& \left. + \lambda_j \bar{\lambda}_j \lambda_i^2 \left(\delta_{(a,b)(c_i, \sigma^{-1}(\overline{c}_j))} \delta_{(b,c)(\sigma^{-1}(\overline{c}_j), c_j))} + \delta_{(a,b)(c_i, c_j)} \delta_{(b,c)(c_j, \sigma^{-1}(\overline{c}_j))} \right) \right. \\
& \left. + \lambda_i \bar{\lambda}_i \bar{\lambda}_j^2 \left(\delta_{(a,b)(c_i, \sigma^{-1}(\overline{c}_i))} \delta_{(b,c)(\sigma^{-1}(\overline{c}_i), \sigma^{-1}(\overline{c}_j))} \right. \right. \\
&\left. \left. + \delta_{(a,b)(c_i, \sigma^{-1}(\overline{c}_j))} \delta_{(b,c)(\sigma^{-1}(\overline{c}_j), \sigma^{-1}(\overline{c}_i))} \right) \right. \\
& \left. + \lambda_j \bar{\lambda}_j \bar{\lambda}_i^2 \left(\delta_{(a,b)(\sigma^{-1}(\overline{c}_i), \sigma^{-1}(\overline{c}_j))} \delta_{(b,c)(\sigma^{-1}(\overline{c}_j), c_j)} \right. \right. \\
& \left. \left. + \delta_{(a,b)( \sigma^{-1}(\overline{c}_i), c_j)} \delta_{(b,c)(c_j, \sigma^{-1}(\overline{c}_j))} \right) \right] \\
& +2 \sum_{(a,b)} \sum_{(c,d)} \delta_{\sigma \tau_{(a,b)(c,d)}} \lambda_i \bar{\lambda}_i \lambda_j \bar{\lambda}_j \\
&\left[ \delta_{(a,b)(c_i, \sigma^{-1}(\overline{c}_j))} \delta_{(c,d)( \sigma^{-1}(\overline{c}_i), c_j)} + \delta_{(a,b)(c_i, c_j)} \delta_{(c,d)(\sigma^{-1}(\overline{c}_i), \sigma^{-1}(\overline{c}_j))} \right] 
\end{split}
\label{A2.doubleexchange}
\end{equation}
\normalsize
\newpage
\begin{itemize}
\item{Crossed double exchange:}
\end{itemize}

\small
\begin{equation}
\begin{split}
[\tau|\bold{T}_i \bold{T}_i \bold{T}_j \bold{T}_j |\sigma \rangle &= N^2 \delta_{\sigma\tau} \bigg[2 \lambda_i \bar{\lambda}_i \lambda_j \bar{\lambda}_j \delta_{c_i \sigma^{-1}(\overline{c}_j)} \delta_{c_j \sigma^{-1}(\overline{c}_i)}  \\
& +\frac{2}{N^2}(\lambda_i- \bar{\lambda}_i) (\lambda_j-\bar{\lambda}_j)(\lambda_i  \bar{\lambda}_j  \delta_{c_i \sigma^{-1}(\overline{c}_j)}  +\bar{\lambda}_i  \lambda_j  \delta_{c_j \sigma^{-1}(\overline{c}_i)}) \\
&+\frac{1}{N^2} \left(\lambda_i^2 \bar{\lambda}_j^2 + \bar{\lambda}_i^2 \lambda_j^2\right) +\frac{1}{N^4} (\lambda_i - \bar{\lambda}_i)^2(\lambda_j- \bar{\lambda}_j)^2 \bigg] \\
& + \sum_{(a,b)} \delta_{\sigma \tau_{(a,b)}} \left[N \lambda_i^2 \lambda_j^2 \delta_{(a,b)(c_i, c_j)} + N \bar{\lambda}_i^2 \bar{\lambda}_j^2 \delta_{(a,b)(\sigma^{-1}(\overline{c}_i), \sigma^{-1}(\overline{c}_j))} \right.\\
& \left. +2N \lambda_i \bar{\lambda}_i \lambda_j \bar{\lambda}_j \left(\delta_{c_i \sigma^{-1}(\overline{c}_j)} \delta_{(a,b)(\sigma^{-1}(\overline{c}_i), c_j)} + \delta_{c_j \sigma^{-1}(\overline{c}_i)} \delta_{(a,b)(c_i, \sigma^{-1}(\overline{c}_j))}\right) \right. \\
& \left. + \frac{2}{N}(\lambda_i - \bar{\lambda}_i) (\lambda_j- \bar{\lambda}_j) \left(\lambda_i  \bar{\lambda}_j \delta_{(a,b)(c_i, \sigma^{-1}(\overline{c}_j))}+ \bar{\lambda}_i  \lambda_j  \delta_{(a,b)(c_j, \sigma^{-1}(\overline{c}_i))} \right. \right. \\
& \left. \left. -\lambda_i  \lambda_j \delta_{(a,b)(c_i, c_j)}- \bar{\lambda}_i \bar{\lambda}_j \delta_{(a,b)(\sigma^{-1}(\overline{c}_i), \sigma^{-1}(\overline{c}_j))} \right) \right] \\
& -\sum_{(a,b)} \sum_{(b,c)} \delta_{\sigma \tau_{(a,b)(b,c)}} \\
&\left[\lambda_j^2 \lambda_i \bar{\lambda}_i \left(\delta_{(a,b)(c_i, c_j)} \delta_{(b,c)(c_j, \sigma^{-1}(\overline{c}_i)))} + \delta_{(a,b)(c_i, \sigma^{-1}(\overline{c}_i))} \delta_{(b,c)(\sigma^{-1}(\overline{c}_i), c_j)} \right) \right. \\
& \left. + \bar{\lambda}_j^2 \lambda_i \bar{\lambda}_i \left(\delta_{(a,b)(c_i, \sigma^{-1}(\overline{c}_i))} \delta_{(b,c)(\sigma^{-1}(\overline{c}_i), \sigma^{-1}(\overline{c}_j))} \right. \right.  \\
&\left. \left. \hspace{1.6cm} + \delta_{(a,b)(c_i, \sigma^{-1}(\overline{c}_j))} \delta_{(b,c)(\sigma^{-1}(\overline{c}_j), \sigma^{-1}(\overline{c}_i))}\right) \right. \\
& \left. +\bar{\lambda}_i^2 \lambda_j \bar{\lambda}_j \left(\delta_{(a,b)(\sigma^{-1}(\overline{c}_i), c_j)} \delta_{(b,c)(c_j, \sigma^{-1}(\overline{c}_j))}\right. \right.  \\
&\left. \left. \hspace{1.6cm} + \delta_{(a,b)(\sigma^{-1}(\overline{c}_i), \sigma^{-1}(\overline{c}_j))} \delta_{(b,c)(\sigma^{-1}(\overline{c}_j), c_j)}\right) \right. \\
& \left. + \lambda_i^2 \lambda_j \bar{\lambda}_j \left(\delta_{(a,b)(c_i, \sigma^{-1}(\overline{c}_j))} \delta_{(b,c)(\sigma^{-1}(\overline{c}_j), c_j)} + \delta_{(a,b)(c_i, c_j)} \delta_{(b,c)(c_j, \sigma^{-1}(\overline{c}_j))}\right) \right] \\
&+2 \sum_{(a,b)} \sum_{(c,d)} \delta_{\sigma \tau_{(a,b)(c,d)}} \lambda_i \bar{\lambda}_i \lambda_j \bar{\lambda}_j \left[\delta_{(a,b)(c_i, \sigma^{-1}(\overline{c}_j))} \delta_{(c,d)(c_j, \sigma^{-1}(\overline{c}_i))} \right. \\
& \left. + \delta_{(a,b)(c_i, c_j)} \delta_{(c,d)(\sigma^{-1}(\overline{c}_i), \sigma^{-1}(\overline{c}_j))}\right] 
\end{split}
\label{A2.crossed}
\end{equation}
\normalsize
\newpage
\begin{itemize}
\item{Triple gluon vertex connecting three hard lines:}
\end{itemize}

\small
\begin{equation}
\begin{split}
[\tau| \bold{T}_g \bold{T}_l \bold{T}_i \bold{T}_j |\sigma \rangle &= \sqrt{T_R} N^2 \delta_{\sigma \tau} \bigg[\frac{1}{N^2} \left((\lambda_l -\bar{\lambda}_l) \left(\lambda_i \bar{\lambda}_j  \delta_{c_i \sigma^{-1}(\overline{c}_j)}- \lambda_j \bar{\lambda}_i \delta_{c_j \sigma^{-1}(\overline{c}_i)}\right) \right.  \\
& \left. +(\lambda_j -\bar{\lambda}_j) \left(\bar{\lambda}_i \lambda_l  \delta_{c_l \sigma^{-1}(\overline{c}_i)}- \lambda_i \bar{\lambda}_l \delta_{c_i \sigma^{-1}(\overline{c}_l)}\right)  \right. \\
&\left.  +(\lambda_i -\bar{\lambda}_i) \left(\bar{\lambda}_l \lambda_j  \delta_{c_j \sigma^{-1}(\overline{c}_l)}-\bar{\lambda}_j \lambda_l \delta_{c_l \sigma^{-1}(\overline{c}_j)}\right)\right) \bigg]\\
&+ \sum_{(a,b)} \delta_{\sigma \tau_{(a,b)}} N \sqrt{T_R} \left[ \lambda_i \bar{\lambda}_j \lambda_l \delta_{(a,b)(c_i, c_l)} \left(\delta_{c_l \sigma^{-1}(\overline{c}_j)} - \delta_{c_i \sigma^{-1}(\overline{c}_j)}\right) \right. \\
&\left. + \bar{\lambda}_i \lambda_j \lambda_l \delta_{(a,b)(c_j, c_l)} \left(\delta_{c_j \sigma^{-1}(\overline{c}_i)} - \delta_{c_l \sigma^{-1}(\overline{c}_i)} \right) \right.  \\
&\left. + \lambda_i \lambda_j \bar{\lambda}_l \delta_{(a,b)(c_i, c_j)} \left( \delta_{c_i \sigma^{-1}(\overline{c}_l)}- \delta_{c_j \sigma^{-1}(\overline{c}_l)} \right) \right. \\
& \left. + \lambda_i \bar{\lambda}_j \bar{\lambda}_l \left( \delta_{(a,b)(c_i, \sigma^{-1}(\overline{c}_l))} \delta_{c_i \sigma^{-1}(\overline{c}_j)} - \delta_{(a,b)(c_i, \sigma^{-1}(\overline{c}_j))} \delta_{c_i \sigma^{-1}(\overline{c}_l)} \right) \right. \\
& \left. + \bar{\lambda}_i \lambda_j \bar{\lambda}_l \left( \delta_{(a,b)(\sigma^{-1}(\overline{c}_i), c_j)} \delta_{c_j \sigma^{-1}(\overline{c}_l)} - \delta_{(a,b)(c_j, \sigma^{-1}(\overline{c}_l))} \delta_{c_j \sigma^{-1}(\overline{c}_i)} \right) \right. \\
& \left. + \bar{\lambda}_i \bar{\lambda}_j \lambda_l \left(\delta_{(a,b)(\sigma^{-1}(\overline{c}_j), c_l)} \delta_{c_l \sigma^{-1}(\overline{c}_i)} - \delta_{(a,b)(\sigma^{-1}(\overline{c}_i), c_l)} \delta_{c_l \sigma^{-1}(\overline{c}_j)} \right)\right] \\
&+ \sum_{(a,b)} \sum_{(b,c)} \delta_{\sigma \tau_{(a,b)(b,c)}} \sqrt{T_R} \left[\lambda_i \lambda_j \delta_{(a,b)(c_i, c_j)} \left(\lambda_l \delta_{(b,c)(c_j, c_l)} - \bar{\lambda}_l \delta_{(b,c)(c_j, \sigma^{-1}(\overline{c}_l))} \right) \right. \\
& \left. - \lambda_i \lambda_l \delta_{(a,b)(c_i, c_l)} \left( \lambda_j \delta_{(b,c)(c_l, c_j)} - \bar{\lambda}_j \delta_{(b,c)(c_l, \sigma^{-1}(\overline{c}_j))} \right) \right. \\
& \left. - \lambda_i \bar{\lambda}_j \delta_{(a,b)(c_i, \sigma^{-1}(\overline{c}_j))} \left(\lambda_l \delta_{(b,c)(\sigma^{-1}(\overline{c}_j), c_l)} - \bar{\lambda}_l \delta_{(b,c)(\sigma^{-1}(\overline{c}_j), \sigma^{-1}(\overline{c}_l))} \right) \right. \\
& \left. + \bar{\lambda}_i \lambda_l \delta_{(a,b)(\sigma^{-1}(\overline{c}_i), c_l)} \left(\lambda_j \delta_{(b,c)(c_l, c_j)} -\bar{\lambda}_j \delta_{(b,c)(c_l, \sigma^{-1}(\overline{c}_j))} \right) \right. \\
& \left. - \bar{\lambda}_i \lambda_j \delta_{(a,b)(\sigma^{-1}(\overline{c}_i), c_j)} \left( \lambda_l \delta_{(b,c)(c_j, c_l)} - \bar{\lambda}_l \delta_{(b,c)(c_j, \sigma^{-1}(\overline{c}_l))} \right) \right. \\
& \left.  - \bar{\lambda}_i \bar{\lambda}_l \delta_{(a,b)(\sigma^{-1}(\overline{c}_i), \sigma^{-1}(\overline{c}_l))} \left(\lambda_j \delta_{(b,c)(\sigma^{-1}(\overline{c}_l), c_j)}- \bar{\lambda}_j \delta_{(b,c)(\sigma^{-1}(\overline{c}_l), \sigma^{-1}(\overline{c}_j))} \right) \right. \\
& \left. +\lambda_i \bar{\lambda}_l \delta_{(a,b)(c_i, \sigma^{-1}(\overline{c}_l))} \left( \lambda_j \delta_{(b,c)(\sigma^{-1}(\overline{c}_l), c_j)} - \bar{\lambda}_j \delta_{(b,c)(\sigma^{-1}(\overline{c}_l), \sigma^{-1}(\overline{c}_j))} \right)\right. \\
& \left. + \bar{\lambda}_i \bar{\lambda}_j \delta_{(a,b)(\sigma^{-1}(\overline{c}_i), \sigma^{-1}(\overline{c}_j))} \left(\lambda_l \delta_{(b,c)(\sigma^{-1}(\overline{c}_j), c_l)} - \bar{\lambda}_l \delta_{(b,c)(\sigma^{-1}(\overline{c}_j), \sigma^{-1}(\overline{c}_l))} \right) \right] 
\end{split}
\label{A2.3legtriple}
\end{equation}
\normalsize
\newpage

\begin{itemize}
\item{Double gluon exchange between three hard lines:}
\end{itemize}

\small
\begin{equation}
\begin{split}
[\tau|(\bold{T}_i \cdot \bold{T}_l) (\bold{T}_i \cdot \bold{T}_j) |\sigma \rangle& =N^2 \delta_{\sigma \tau} \\
&\bigg[ - \lambda_i \bar{\lambda}_i \left(\lambda_j \bar{\lambda}_l \delta_{c_i \sigma^{-1}(\overline{c}_l)} \delta_{c_j \sigma^{-1}(\overline{c}_i)} +  \bar{\lambda}_j \lambda_l \delta_{c_i \sigma^{-1}(\overline{c}_j)} \delta_{c_l \sigma^{-1}(\overline{c}_i)} \right) \\
&+ \frac{1}{N^2} (\lambda_i - \bar{\lambda}_i)(\lambda_l - \bar{\lambda}_l) \left(\lambda_i \bar{\lambda}_j \delta_{c_i \sigma^{-1}(\overline{c}_j)} + \bar{\lambda}_i \lambda_j \delta_{c_j \sigma^{-1}(\overline{c}_i)} \right) \\
& -\frac{1}{N^2} \left(\lambda_i^2 \bar{\lambda}_j \lambda_l \delta_{c_l \sigma^{-1}(\overline{c}_j)} + \bar{\lambda}_i^2 \lambda_j \bar{\lambda}_l \delta_{c_j \sigma^{-1}(\overline{c}_l)}\right)  \\
&+ \frac{1}{N^2} \lambda_i \bar{\lambda}_i (\lambda_j - \bar{\lambda}_j) \left(\bar{\lambda}_l \delta_{c_i \sigma^{-1}(\overline{c}_l)} -\lambda_l \delta_{c_l \sigma^{-1}(\overline{c}_i)} \right)  \\
& + \frac{1}{N^4} (\lambda_i - \bar{\lambda}_i)^2 (\lambda_j - \bar{\lambda}_j) (\lambda_l - \bar{\lambda}_l)   \bigg] \\
&+ \sum_{(ab)} \delta_{\sigma \tau_{(a,b)}} \\
&\left\{ N \lambda_i \bar{\lambda}_i \left[\lambda_j \delta_{c_j \sigma^{-1}(\overline{c}_i)}\left(\lambda_l \delta_{(a,b)(c_i, c_l)} - \bar{\lambda}_l \delta_{(a,b)(c_i, \sigma^{-1}(\overline{c}_l))} \right) \right. \right. \\
&\left. \left. + \lambda_l \delta_{c_l \sigma^{-1}(\overline{c}_i)} \left( \lambda_j \delta_{(a,b)(c_i, c_j)} -\bar{\lambda}_j \delta_{(a,b)(c_i, \sigma^{-1}(\overline{c}_j))} \right)  \right. \right. \\
&\left. \left. -\bar{\lambda}_l \delta_{c_i \sigma^{-1}(\overline{c}_l)} \left(\lambda_j \delta_{(a,b)(\sigma^{-1}(\overline{c}_i), c_j)}- \bar{\lambda}_j \delta_{(a,b)(\sigma^{-1}(\overline{c}_i), \sigma^{-1}(\overline{c}_j))} \right) \right. \right. \\
&\left. \left. - \bar{\lambda}_j \delta_{c_i \sigma^{-1}(\overline{c}_j)} \left(\lambda_l \delta_{(a,b)(\sigma^{-1}(\overline{c}_i), c_l)}- \bar{\lambda}_l \delta_{(a,b)(\sigma^{-1}(\overline{c}_i), \sigma^{-1}(\overline{c}_l))} \right)  \right] \right. \\
&\left. +N \left[ - \lambda_i^2 \bar{\lambda}_j \delta_{c_i \sigma^{-1}(\overline{c}_j)} \left(\lambda_l \delta_{(a,b)(c_i, c_l)} - \bar{\lambda}_l \delta_{(a,b)(c_i, \sigma^{-1}(\overline{c}_l))} \right) \right. \right. \\
&\left. \left. + \bar{\lambda}_i^2 \lambda_j \delta_{c_j \sigma^{-1}(\overline{c}_i)} \left(\lambda_l \delta_{(a,b)(c_j, c_l)} - \bar{\lambda}_l \delta_{(a,b)(c_j, \sigma^{-1}(\overline{c}_l))} \right) \right. \right. \\
&\left. \left. - \bar{\lambda}_i^2 \bar{\lambda}_j \lambda_l \delta_{c_l \sigma^{-1}(\overline{c}_j)} \delta_{(a,b)(\sigma^{-1}(\overline{c}_i), c_l)} - \lambda_i^2 \lambda_j \bar{\lambda}_l \delta_{c_j \sigma^{-1}(\overline{c}_i)} \delta_{(a,b)(c_i, c_j)} \right] \right. \\
&\left. + \frac{1}{N} (\lambda_i - \bar{\lambda}_i) \Big[(\lambda_j - \bar{\lambda}_j)\left(\lambda_i \bar{\lambda}_l \delta_{(a,b)(c_i, \sigma^{-1}(\overline{c}_l))}- \lambda_i  \lambda_l \delta_{(a,b)(c_i, c_l)} \right. \right. \\
&\left. \left. + \bar{\lambda}_i \lambda_l \delta_{(a,b)(\sigma^{-1}(\overline{c}_i), c_l)}- \bar{\lambda}_i \bar{\lambda}_l \delta_{(a,b)(\sigma^{-1}(\overline{c}_i), \sigma^{-1}(\overline{c}_l))} \right)\right.  \\
&\left.  +(\lambda_l - \bar{\lambda}_l) \left(\lambda_i \bar{\lambda}_j  \delta_{(a,b)(c_i, \sigma^{-1}(\overline{c}_j))}+ \bar{\lambda}_i \lambda_j \delta_{(a,b)(\sigma^{-1}(\overline{c}_i), c_j)} \right. \right. \\
& \left. \left. - \lambda_i \lambda_j  \delta_{(a,b)(c_i, c_j)}  - \bar{\lambda}_i  \bar{\lambda}_j \delta_{(a,b)(\sigma^{-1}(\overline{c}_i), \sigma^{-1}(\overline{c}_j))}\right) \Big] \right\} \\
& + \sum_{(a,b)} \sum_{(b,c)} \delta_{\sigma \tau_{(a,b)(b,c)}}\\
& \left\{ \lambda_i^2 \lambda_j \delta_{(a,b)(c_i, c_j)} \left( \lambda_l \delta_{(b,c) (c_j, c_l)} - \bar{\lambda}_l \delta_{(b,c)(c_j, \sigma^{-1}(\overline{c}_l))} \right) \right. \\
& \left. + \bar{\lambda}_i^2 \lambda_l \delta_{(a,b)(\sigma^{-1}(\overline{c}_i), c_l)} \left( \lambda_j \delta_{(b,c)(c_l, c_j)} - \bar{\lambda}_j \delta_{(b,c)(c_l, \sigma^{-1}(\overline{c}_j))} \right) \right. \\
& \left. - \bar{\lambda}_i^2 \bar{\lambda}_l \delta_{(a,b)(\sigma^{-1}(\overline{c}_i), \sigma^{-1}(\overline{c}_l))}\hspace{-0.1cm} \left(\lambda_j \delta_{(b,c)(\sigma^{-1}(\overline{c}_l), c_j)} - \bar{\lambda}_j \delta_{(b,c)(\sigma^{-1}(\overline{c}_l), \sigma^{-1}(\overline{c}_j))} \hspace{-0.1cm}\right) \right. \\
& \left.  - \lambda_i^2 \bar{\lambda}_j \delta_{(a,b)(c_i, \sigma^{-1}(\overline{c}_j))} \left(\lambda_l \delta_{(b,c)(\sigma^{-1}(\overline{c}_j), c_l)} - \bar{\lambda}_l \delta_{(b,c)(\sigma^{-1}(\overline{c}_j), \sigma^{-1}(\overline{c}_l))} \right)\right\} \\
& +\sum_{(a,b)} \sum_{(c,d)} \delta_{\sigma \tau_{(a,b)(c,d)}} \lambda_i \bar{\lambda}_i \\
 &\left[  \lambda_j \delta_{(a,b)(c_i, c_j)} \left( \lambda_l \delta_{(c,d)(\sigma^{-1}(\overline{c}_i), c_l)} - \bar{\lambda}_l \delta_{(c,d)(\sigma^{-1}(\overline{c}_i), \sigma^{-1}(\overline{c}_l))} \right) \right. \\
& \left. - \bar{\lambda}_j \delta_{(a,b)(c_i, \sigma^{-1}(\overline{c}_j))} \left(\lambda_l \delta_{(c,d)(\sigma^{-1}(\overline{c}_i), c_l)} - \bar{\lambda}_l \delta_{(c,d)(\sigma^{-1}(\overline{c}_i), \sigma^{-1}(\overline{c}_l))} \right) \right. \\
& \left. + \lambda_l \delta_{(a,b)(c_i, c_l)} \left(\lambda_j \delta_{(c,d)(\sigma^{-1}(\overline{c}_i), c_j)} - \bar{\lambda}_j \delta_{(c,d)(\sigma^{-1}(\overline{c}_i), \sigma^{-1}(\overline{c}_j))} \right) \right. \\
& \left. - \bar{\lambda}_l \delta_{(a,b)(c_i, \sigma^{-1}(\overline{c}_l))} \left(\lambda_j \delta_{(c,d)(\sigma^{-1}(\overline{c}_i), c_j)} - \bar{\lambda}_j \delta_{(c,d)(\sigma^{-1}(\overline{c}_i), \sigma^{-1}(\overline{c}_j))}\right) \right]
\end{split}
\label{A2.3leg2}
\end{equation}
\normalsize
%%%%%%%%%%%%
\section{Diagrams involving the gluon self energy}
\label{sec:SelfEnergy}
The gluon self energy diagrams at two-loop level can be expressed in the following form after reduction (neglecting terms which vanish due to colour conservation)
\begin{equation}
\begin{split}
\begin{gathered}
\vspace{-0.2cm}
\includegraphics[width=2.2cm]{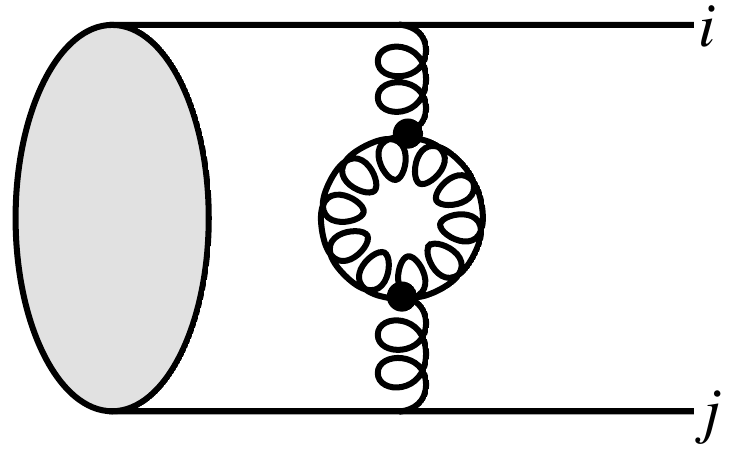}
\end{gathered}
+
\begin{gathered}
\vspace{-0.2cm}
\includegraphics[width=2.15cm]{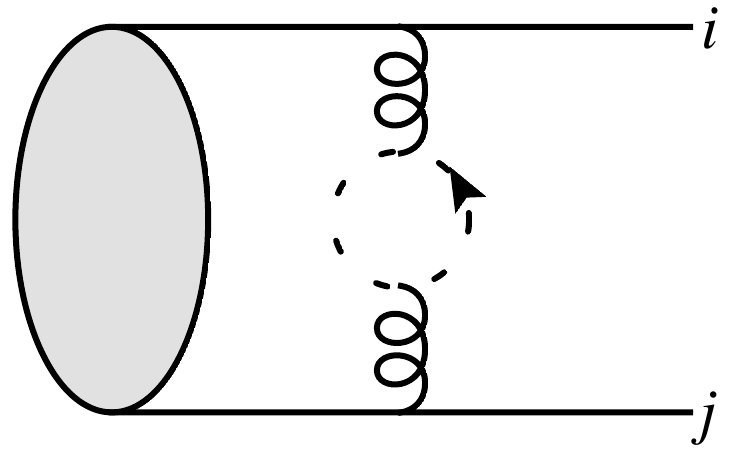}
\end{gathered}
=&\frac{\alpha_s^2}{2} \sum_{i,j} T_R (\bold{T}_i\cdot \bold{T}_j)\, 2C_A \\
&\left[(4-2d) I_5^{(ij)} -2(p_i\cdot p_j) I_3^{(ij)} -(p_i\cdot p_j) I_6^{(ij)} \right] \ ,
\end{split}
\end{equation}
and 
\begin{equation}
\begin{split}
\begin{gathered}
\vspace{-0.2cm}
\includegraphics[width=2.2cm]{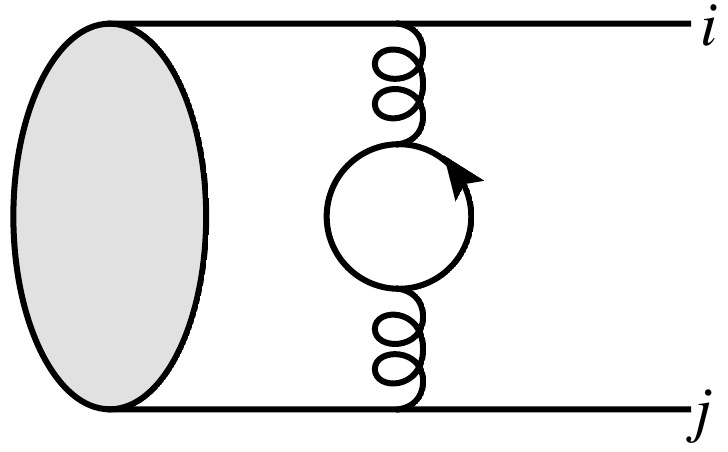}
\end{gathered}
=&\frac{\alpha_s^2}{2} \sum_{i,j} 4 T_R (\bold{T}_i\cdot \bold{T}_j)\, n_f \left[2 I_5^{(ij)} -\frac{1}{2} (p_i\cdot p_j) I_3^{(ij)}-(p_i\cdot p_j) I_6^{(ij)} \right] \ .
\end{split}
\end{equation}
Again, we have used unordered colour sums. This requires the inclusion of a factor of $1/2$ in order to avoid overcounting for the bubble diagrams. \\
The loop integrals pertaining to the diagrams above are given by
\begin{equation}
\begin{split}
&I^{(ij)}_5=\mu^{4\epsilon} \int_{k,q} \frac{(p_i\cdot k)(p_j\cdot k)}{[p_i\cdot q+i0][-p_j\cdot q+i0][k^2+i0][(q-k)^2+i0][q^2+i0]^2} \ , \\
&I^{(ij)}_6=\mu^{4\epsilon} \int_{k,q} \frac{1}{[p_i\cdot q+i0][-p_j\cdot q+i0][k^2+i0][q^2+i0]^2} \ , \\
&I^{(ij)}_3=\mu^{4\epsilon} \int_{k,q} \frac{1}{[p_i\cdot q+i0][-p_j\cdot q+i0][k^2+i0][(q-k)^2+i0][q^2+i0]}\ ,
\end{split}
\end{equation}
where we have used the short-hand notation $\int_{k,q}\equiv \int \frac{\mathrm{\dbar}^dk}{i\pi^{d/2}} \int \frac{\mathrm{\dbar}^dq}{i\pi^{d/2}}$. Furthermore note that according to our definition the $i0$-prescription of an Eikonal propagator with momentum $p_i$ is given by $i0 (T\cdot p_i)^2$ and for the soft gluon propagator of momentum $q$ we have $i0 (T\cdot q)^2$. To avoid clutter we do not explicitly write these factors for the $i0$-prescription in the following. However, they can be easily reinstated with the rule above. \\
The contribution $I^{(ij)}_3$ does not contain a propagator to second power anymore. Due to its propagator structure this contribution is treated in App.~\ref{app:FTT2}, Eq. (\ref{2ldiagram_3}). For the contributions with higher order propagators we find
\begin{equation}
\begin{split}
I^{(ij)}_5=\mu^{4\epsilon} \int_{k,q} (p_i\cdot k)(p_j\cdot k) &\left\{-\frac{(2\pi i)^2 \tilde{\delta}(q) \tilde{\delta}(k)}{[p_i\cdot(k+q)+i0][-p_j\cdot(k+q)+i0][(k+q)^2+i0]^2}\right. \\
& \left. + \frac{2 (2\pi i)^2 \tilde{\delta}(k)\delta(p_i\cdot q)}{[-p_j\cdot q+i0][(q-k)^2+i0][q^2+i0]^2}\right. \\
&\left. +\frac{2(2\pi i)^2 (-\delta'(q^2)\theta(q^0)) \tilde{\delta}(k)}{[p_i\cdot q+i0][-p_j\cdot q+i0][(q-k)^2+i0]}\right. \\
&\left. +\frac{2(2\pi i)^3 \tilde{\delta}(k) \tilde{\delta}(q-k) \delta(p_i\cdot q)}{[-p_j\cdot q+i0][q^2+i0]^2}\right. \\
&\left. +\frac{2(2\pi i)^3 (-\delta'(q^2) \theta(q^0)) \tilde{\delta}(k) \delta(p_i\cdot q)}{[-p_j\cdot q+i0][(q-k)^2+i0]}\right. \\
&\left. + \frac{2(2\pi i)^3 (-\delta'(q^2) \theta(q^0)) \tilde{\delta}(k) \tilde{\delta}(q-k)}{[p_i\cdot q+i0][-p_j\cdot q+i0]}\right. \\
&\left. + \frac{2(2\pi i)^4 (-\delta'(q^2) \theta(q^0)) \tilde{\delta}(k) \tilde{\delta}(q-k) \delta(p_i\cdot q)}{[-p_j\cdot q+i0]}\right\} \ ,
\end{split}
\label{2ldiagram_5}
\end{equation}

\begin{equation}
\begin{split}
I^{(ij)}_6=\mu^{4\epsilon}\int_{k,q} &\left\{\frac{(2\pi i)^2 \tilde{\delta}(k) \delta(p_i\cdot q)}{[-p_j\cdot q+i0][q^2+i0]^2}+\frac{(2\pi i)^2 (- \delta'(q^2) \theta(q^0)) \tilde{\delta}(k)}{[p_i\cdot q+i0][-p_j\cdot q+i0]}\right. \\
& \left. + \frac{(2\pi i)^3 (-\delta'(q^2) \theta(q^0)) \tilde{\delta}(k) \delta(p_i\cdot q)}{[-p_j\cdot q+i0]}\right\} \ .
\end{split}
\label{2ldiagram_6}
\end{equation}
\newpage
\section{List of integrals}
\label{sec:Integrals}
\subsection{One loop and one emission}\label{app:FTT11}
Performing a Passarino-Veltman reduction one obtains the following contributions for the diagrams at one-loop, one-emission order 
\begin{equation}
\begin{gathered}
\hspace{0.2cm}
\vspace{0.1cm}
\includegraphics[width=2.2cm]{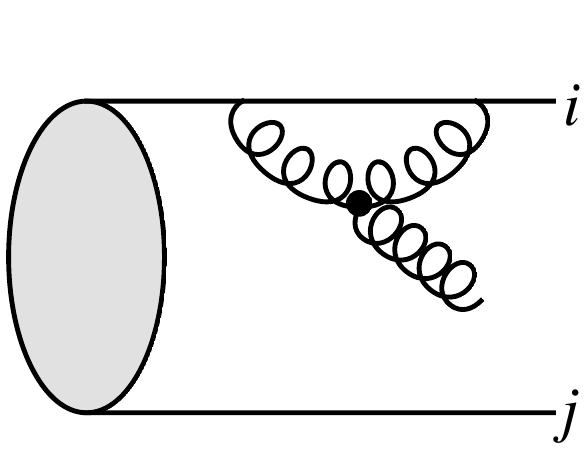}
\end{gathered}
=\alpha_s \, g \sum_{i,j} T_R \bold{T}_i^a C_A  (p_i\cdot \varepsilon^*(q)) \left[I_{10}^{(ij)}- \frac{2}{p_i\cdot q} I_0^{(ij)}\right] \ ,
\end{equation}
\begin{equation}
\begin{gathered}
\vspace{-0.2cm}
\includegraphics[width=2.2cm]{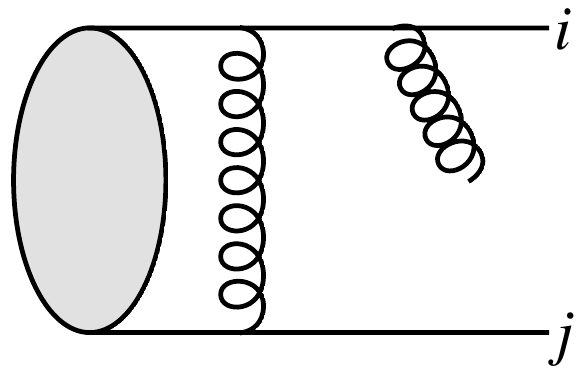}
\end{gathered}
=\alpha_s \, g \sum_{i,j} \bold{T}_i^a (\bold{T}_i\cdot \bold{T}_j) (-i) \frac{p_i\cdot p_j}{p_i\cdot q} (p_i\cdot \varepsilon^*(q)) I_7^{(ij)} \ ,
\end{equation}
\begin{equation}
\begin{gathered}
\hspace{0.1cm}
\vspace{-0.2cm}
\includegraphics[width=2.2cm]{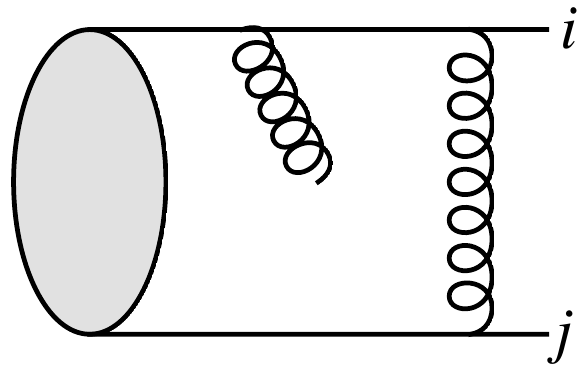}
\end{gathered}
=\alpha_s \, g \sum_{i,j} (\bold{T}_i\cdot \bold{T}_j) \bold{T}_i^a (-i) (p_i\cdot p_j) (p_i\cdot \varepsilon^*(q)) I_8^{(ij)}\ ,
\end{equation}
\begin{equation}
\begin{split}
\begin{gathered}
\vspace{-0.2cm}
\includegraphics[width=2.2cm]{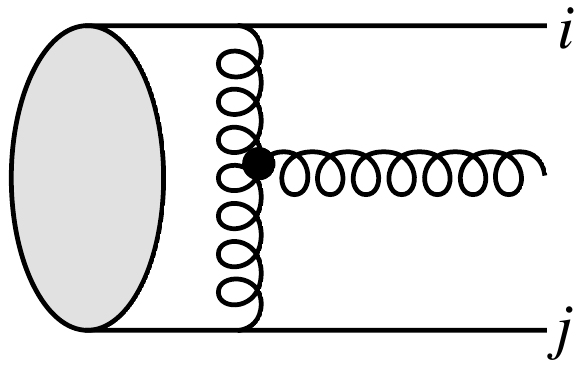}
\end{gathered}
=&-\frac{\alpha_s \, g}{2}\sum_{i,j} f^{abc} \bold{T}_i^b \bold{T}_j^c \Bigg\{(p_j\cdot q)(p_i\cdot \varepsilon^*(q)) I_9^{(ij)} \\
& - \frac{p_i\cdot q}{p_j\cdot q} (p_j\cdot \varepsilon^*(q)) I_{10}^{(ij)} -\frac{p_i\cdot p_j}{2}  \left[\frac{p_i\cdot \varepsilon^*(q)}{p_i\cdot q}+ \frac{p_j\cdot \varepsilon^*(q)}{p_j\cdot q}\right] I_7^{(ij)}\Bigg\} \ ,
\end{split}
\end{equation}
The integrals in the equations above are given by 
\begin{equation}
\begin{aligned}
&I^{(ij)}_7= \mu^{2\epsilon} \int_k  \frac{1}{[p_i\cdot(k+q)+i0][-p_j\cdot k+i0][k^2+i0]},  \\
&I^{(ij)}_8= \mu^{2\epsilon} \int_k  \frac{1}{[p_i\cdot (k+q)+i0][p_i\cdot k+i0][-p_j\cdot k+i0][k^2+i0]}, \\
& I^{(ij)}_9=\mu^{2\epsilon} \int_k  \frac{1}{[p_i\cdot(k+q)+i0][-p_j\cdot k+i0][k^2+i0][(k+q)^2+i0]},  \\
&I^{(ij)}_{10}=  \mu^{2\epsilon} \int_k  \frac{1}{[-p_i\cdot k+i0][k^2+i0][(k+q)^2+i0]},  \\
&I^{(ij)}_0= \mu^{2\epsilon} \int_k \frac{1}{[(k+q)^2+i0][k^2+i0]} \ .
\end{aligned}
\label{(1,1)integrals}
\end{equation}
Using the Feynman tree theorem and filtering out contributions which give zero, such as $\tilde{\delta}(k) \delta(p_i\cdot(k+q))$ and $\tilde{\delta}(k) \delta(p_i\cdot(k+q)) \delta(k\cdot q)$ - for $q$ being on-shell the conditions for $E_k$ lead to a contradiction - results in
\begin{equation}
\begin{split}
I^{(ij)}_7=-\mu^{2\epsilon} \int_k & \left\{\frac{(2\pi i) \tilde{\delta}(k)}{[p_i\cdot(k+q)+i0][-p_j\cdot k+i0]} +\frac{(2\pi i)\delta(p_i\cdot k)}{[(k-q)^2+i0][p_j\cdot (q-k)+i0]} \right. \\
&\left.+ \frac{(2 \pi i)^2\tilde{\delta}(k) \delta(p_i\cdot (k+q))}{[-p_j\cdot k+i0]}\right\} \ ,
\end{split}
\label{(vi)}
\end{equation}
\begin{equation}
\begin{split}
I^{(ij)}_8= -\mu^{2\epsilon} \int_k &\left\{\frac{(2\pi i)\tilde{\delta}(k)}{[p_i\cdot(k+q)+i0][p_i\cdot k+i0][-p_j\cdot k+i0]} \right. \\
& \left. +\frac{(2\pi i) \delta(p_i\cdot k)}{[k^2+i0][p_i\cdot (k+q)+i0][-p_j\cdot k+i0]} +\frac{(2\pi i)^2 \tilde{\delta}(k) \delta(p_i\cdot k)}{[p_i\cdot(k+q)+i0][-p_j\cdot k]} \right. \\
& \left. +\frac{(2\pi i) \delta(p_i\cdot k)}{[(k-q)^2+i0][p_i\cdot(k-q)+i0][p_j\cdot(q-k)+i0]}  \right\} \ ,
\end{split}
\label{(viii)}
\end{equation}
\begin{equation}
\begin{split}
I^{(ij)}_9= - \mu^{2\epsilon} \int_k &\left\{ \frac{(2\pi i) \tilde{\delta}(k)}{[(k+q)^2+i0][p_i\cdot(k+q)+i0][-p_j\cdot k+i0]}\right.  \\
&\left. +\frac{(2\pi i)\tilde{\delta}(k)}{[(k-q)^2+i0][p_i\cdot k+i0][p_j\cdot (q-k)+i0]}\right. \\ 
& \left. +\frac{(2\pi i) \delta(p_i\cdot k)}{[(k-q)^2+i0][k^2+i0][p_j\cdot(q-k)+i0]}\right. \\
&\left. + \frac{(2\pi i)^2 \tilde{\delta}(k) \tilde{\delta}(k+q)}{[p_i\cdot(k+q)+i0][-p_j\cdot k+i0]}+ \frac{(2\pi i)^2 \tilde{\delta}(k) \delta(p_i\cdot k)}{[(k-q)^2+i0][p_j\cdot (q-k)+i0]} \right\} \ ,
\end{split}
\label{(iv)}
\end{equation}
\begin{equation}
\begin{split}
I^{(ij)}_{10}= - \mu^{2\epsilon} \int_k &\left\{\frac{(2\pi i) \tilde{\delta}(k)}{[(k+q)^2+i0][-p_i\cdot k+i0]}+ \frac{(2\pi i) \tilde{\delta}(k)}{[(k-q)^2+i0][p_i\cdot (q-k)+i0]} \right. \\
& \left.+ \frac{(2\pi i)\tilde{\delta}(k) \tilde{\delta}(k+q)}{[-p_i\cdot k+i0]} \right\} \ ,
\end{split}
\label{(v)}
\end{equation}
\subsection{Two loops}\label{app:FTT2}
The amplitudes of the Feynman diagrams involving two hard lines at two-loop level are given by
\begin{equation}
\begin{gathered}
\vspace{-0.2cm}
\includegraphics[width=2.1cm]{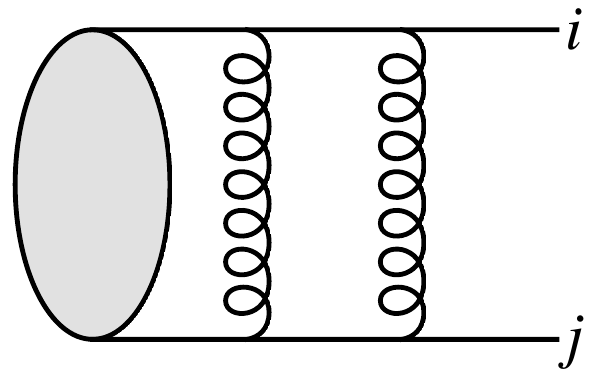}
\end{gathered}
=-\frac{\alpha_s^2}{2} \sum_{i,j} (\bold{T}_i\cdot \bold{T}_j) (\bold{T}_i\cdot \bold{T}_j) (p_i\cdot p_j)^2\,  I_1^{(ij)} \ ,
\end{equation}
\begin{equation}
\begin{gathered}
\vspace{-0.2cm}
\includegraphics[width=2.2cm]{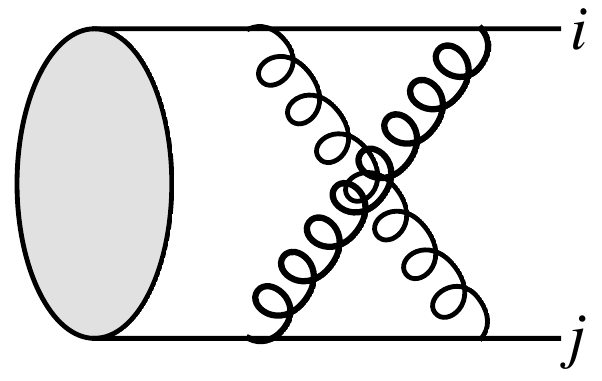}
\end{gathered}
=-\frac{\alpha_s^2}{2} \sum_{i,j} \bold{T}_i^b \bold{T}_i^a \bold{T}_j^a \bold{T}_j^b (p_i\cdot p_j)^2 \, I_2^{(ij)}\ ,
\end{equation}
\begin{equation}
\begin{gathered}
\vspace{-0.2cm}
\includegraphics[width=2.2cm]{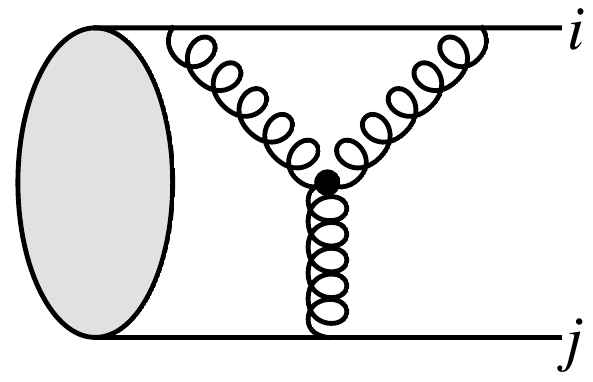}
\end{gathered}
=-\alpha_s^2 \sum_{i,j} T_R (\bold{T}_i\cdot \bold{T}_j) C_A (p_i\cdot p_j)\left[ I_4^{(ij)} -2 I_3^{(ij)}\right] \ .
\end{equation}
The two-loop integrals pertaining to the amplitudes above have the following form
\begin{equation}
\begin{aligned}
& I^{(ij)}_1=\mu^{4\epsilon} \int_{k,q} \frac{1}{[p_i\cdot (k+q)+i0][-p_j\cdot(k+q)+i0][k^2+i0][q^2+i0][p_i\cdot q+i0][-p_j\cdot q+i0]} \\
& I^{(ij)}_2= \, \mu^{4\epsilon} \int_{k,q} \frac{1}{[p_i\cdot(k+q)+i0][-p_j\cdot(k+q)+i0][k^2+i0][q^2+i0][p_i\cdot q+i0][-p_j\cdot k+i0]} \\
& I^{(ij)}_3=\mu^{4\epsilon} \int_{k,q} \frac{1}{[p_i\cdot k+i0][-p_j\cdot k+i0][k^2+i0][q^2+i0][(q-k)^2+i0]} \\
& I^{(ij)}_4=\mu^{4\epsilon} \int_{k,q} \frac{1}{[p_i\cdot k+i0][-p_j\cdot(k+q)+i0][k^2+i0][q^2+i0][(k+q)^2+i0]}  \ .
\end{aligned}
\label{2lintegrals2leg}
\end{equation}
\raggedbottom
Applying the FTT to each integral of the equation above leads us to 
\begin{equation}
\begin{split}
I^{(ij)}_1=\mu^{4\epsilon} \int_{k,q} & \left\{\frac{(2\pi i)^2 \tilde{\delta}(q)\tilde{\delta}(k)}{[p_i\cdot(k+q)+i0][-p_j\cdot(k+q)+i0][p_i\cdot q+i0][-p_j\cdot q+i0]} \right. \\
& \left. + \frac{(2\pi i)^2 \tilde{\delta}(k) \delta(p_i\cdot q)}{[p_i\cdot(k+q)+i0][-p_j\cdot (k+q)+i0][q^2+i0][-p_j\cdot q+i0]} \right. \\
&\left. + \frac{(2\pi i)^2 \delta(p_i\cdot k) \delta(p_i\cdot q)}{[-p_j\cdot k+i0][q^2+i0][(k-q)^2+i0][-p_j\cdot q+i0]} \right. \\
&\left. + \frac{(2\pi i)^2 \tilde{\delta}(q)\delta(p_i\cdot k)}{[p_i\cdot q+i0][-p_j\cdot k+i0][(k-q)^2+i0][-p_j\cdot q+i0]}\right. \\
& \left. - \frac{(2\pi i)^2 \tilde{\delta}(k) \delta(p_i\cdot q)}{[p_i\cdot (q-k)+i0][-p_j\cdot q+i0][(q-k)^2+i0][-p_j\cdot (q-k)+i0]} \right. \\
&\left. + \frac{(2\pi i)^3 \tilde{\delta}(k-q) \delta(p_i\cdot k) \delta(p_i\cdot q)}{[q^2+i0][-p_j\cdot k+i0][-p_j\cdot q +i0]}\right.\\
&\left. +\frac{(2\pi i)^3 \tilde{\delta}(q) \tilde{\delta}(k-q) \delta(p_i\cdot k)}{[p_i\cdot q+i0][-p_j\cdot q+i0][-p_j\cdot k+i0]}\right\} \ ,
\end{split}
\label{2ldiagram_1}
\end{equation}
\begin{equation}
\begin{split}
I^{(ij)}_2=\mu^{4\epsilon}  \int_{k,q} & \left\{\frac{(2\pi i)^2 \tilde{\delta}(k) \tilde{\delta}(q)}{[p_i\cdot (k+q)+i0][p_i\cdot q+i0][-p_j\cdot(k+q)+i0][-p_j\cdot k+i0]}\right. \\
&\left. +\frac{(2\pi i)^2 \tilde{\delta}(k) \delta(p_i\cdot q)}{[p_i\cdot (k+q)+i0][q^2+i0][-p_j\cdot (k+q)+i0][-p_j\cdot k+i0]} \right. \\
&\left. +\frac{(2\pi i)^2 \, \delta(p_i\cdot k)\delta(p_i\cdot q)}{[(k-q)^2+i0][q^2+i0][p_j\cdot (q-k)+i0][-p_j\cdot k+i0]}\right. \\
&\left. +\frac{(2\pi i)^2 \tilde{\delta}(q) \delta(p_i\cdot k)}{[p_i\cdot q+i0][(k-q)^2+i0][p_j\cdot (q-k)+i0][-p_j\cdot k+i0]}\right. \\
&\left. + \frac{(2\pi i)^2 \, \delta(p_i\cdot k) \delta(p_j\cdot q)}{[p_i\cdot (k+q)+i0][(k+q)^2+i0][q^2+i0][-p_j\cdot k+i0]} \right. \\
&\left. - \frac{(2\pi i)^2 \tilde{\delta}(k) \delta(p_i\cdot q)}{[p_i\cdot (q-k)+i0][(q-k)^2+i0][-p_j\cdot q+i0][-p_j\cdot k+i0]}\right. \\
&\left. + \frac{(2\pi i)^3 \, \delta(p_i\cdot k) \delta(p_i\cdot q) \delta(p_j\cdot (q-k))}{[(k-q)^2+i0][q^2+i0][-p_j\cdot k+i0]}\right. \\
&\left. + \frac{(2\pi i)^3 \, \tilde{\delta}(q) \delta(p_i\cdot k) \delta(p_j\cdot(q-k))}{[p_i\cdot q+i0][(k-q)^2+i0][-p_j\cdot k+i0]}\right. \\
&\left. +\frac{(2\pi i)^3 \tilde{\delta}(k-q) \delta(p_i\cdot k) \delta(p_i\cdot q)}{[q^2+i0][p_j\cdot (q-k)+i0][-p_j\cdot k+i0]} \right. \\
&\left. +\frac{(2\pi i)^3 \tilde{\delta}(k-q) \delta(p_j\cdot (q-k)) \delta(p_i\cdot k)}{[p_i\cdot q+i0][q^2+i0][-p_j\cdot k+i0]} \right. \\
&\left. +\frac{(2\pi i)^3 \tilde{\delta}(q) \tilde{\delta}(k-q) \delta(p_i\cdot k)}{[p_i\cdot q+i0][p_j\cdot (q-k)+i0][-p_j\cdot k+i0]} \right. \\
&\left. + \frac{(2\pi i)^4 \tilde{\delta}(k-q) \delta(p_i\cdot k) \delta(p_i\cdot q) \delta(p_j\cdot (q-k))}{[q^2+i0][-p_j\cdot k+i0]}\right\} \ ,
\end{split}
\label{2ldiagram_2}
\end{equation}
\begin{equation}
\begin{split}
I^{(ij)}_3=\mu^{4\epsilon} \int_{k,q} &\left\{\frac{(2\pi i)^2 \tilde{\delta}(k) \tilde{\delta}(q)}{[p_i\cdot k+i0][(q-k)^2+i0][-p_j\cdot k+i0]} \right. \\
&\left. + \frac{(2\pi i)^2 \tilde{\delta}(q) \delta(p_i\cdot k)}{[(q-k)^2+i0][k^2+i0][-p_j\cdot k+i0]}\right. \\
& \left. +\frac{(2\pi i)^2 \tilde{\delta}(q) \tilde{\delta}(k)}{[p_i\cdot k+i0][(k+q)^2+i0][-p_j\cdot k+i0]}\right. \\
&\left. +\frac{(2\pi i)^2 \tilde{\delta}(q) \delta(p_i\cdot k)}{[(k+q)^2+i0][k^2+i0][-p_j\cdot k+i0]}\right. \\
&\left. -\frac{(2\pi i)^2 \tilde{\delta}(k) \tilde{\delta}(q)}{[p_i\cdot(q-k)+i0][(q-k)^2+i0][-p_j\cdot (q-k)+i0]}\right.\\
& \left. +\frac{(2\pi i)^3 \tilde{\delta}(q) \tilde{\delta}(k) \tilde{\delta}(q-k)}{[p_i\cdot k+i0][-p_j\cdot k+i0]}+\frac{(2\pi i)^3 \tilde{\delta}(q) \tilde{\delta}(q-k) \delta(p_i\cdot k)}{[k^2+i0][-p_j\cdot k+i0]}\right. \\
&\left. +\frac{(2\pi i)^3 \tilde{\delta}(q) \tilde{\delta}(k) \tilde{\delta}(k+q)}{[p_i\cdot k+i0][-p_j\cdot k+i0]}+\frac{(2\pi i)^3 \tilde{\delta}(q) \tilde{\delta}(k+q) \delta(p_i\cdot k)}{[k^2+i0][-p_j\cdot k+i0]}\right\} \ ,
\end{split}
\label{2ldiagram_3}
\end{equation}
\begin{equation}
\begin{split}
I^{(ij)}_4=\mu^{4\epsilon} \int_{k,q} & \left\{\frac{(2\pi i)^2 \tilde{\delta}(k) \tilde{\delta}(q)}{[p_i\cdot k+i0][(k+q)^2+i0][-p_j\cdot(k+q)+i0]} \right. \\
&\left. +\frac{(2\pi i)^2 \tilde{\delta}(q) \delta(p_i\cdot k)}{[k^2+i0][(k+q)^2+i0][-p_j\cdot (k+q)+i0]} \right. \\
&\left. + \frac{(2\pi i)^2 \tilde{\delta}(k) \tilde{\delta}(q)}{[p_i\cdot k+i0][(q-k)^2+i0][-p_j\cdot q+i0]}\right. \\
&\left. + \frac{(2\pi i)^2 \tilde{\delta}(q) \delta(p_i\cdot k)}{[k^2+i0][(q-k)^2+i0][-p_j\cdot q+i0]} \right. \\ 
&\left. - \frac{(2\pi i)^2 \tilde{\delta}(q) \tilde{\delta}(k)}{[p_i\cdot (k-q)+i0][(k-q)^2+i0][-p_j\cdot k+i0]}\right. \\
&\left. +\frac{(2\pi i)^3 \tilde{\delta}(k) \tilde{\delta}(q) \tilde{\delta}(k+q)}{[p_i\cdot k+i0][-p_j\cdot (k+q)+i0]}\right. \\
&\left. + \frac{(2\pi i)^3 \tilde{\delta}(q) \tilde{\delta}(k+q) \delta(p_i\cdot k)}{[k^2+i0][-p_j\cdot (k+q)+i0]}+ \frac{(2\pi i)^3 \tilde{\delta}(q) \tilde{\delta}(k) \tilde{\delta}(q-k)}{[p_i\cdot k+i0][-p_j\cdot q+i0]}\right. \\
& \left.  +\frac{(2\pi i)^3 \tilde{\delta}(q-k) \tilde{\delta}(q) \delta(p_i\cdot k)}{[k^2+i0][-p_j\cdot q+i0]}\right\} \ .
\end{split}
\label{2ldiagram_4}
\end{equation}
The amplitudes of the diagrams involving three hard lines can be expressed as
\begin{equation}
\begin{split}
\begin{gathered}
\vspace{-0.2cm}
\includegraphics[width=2.2cm]{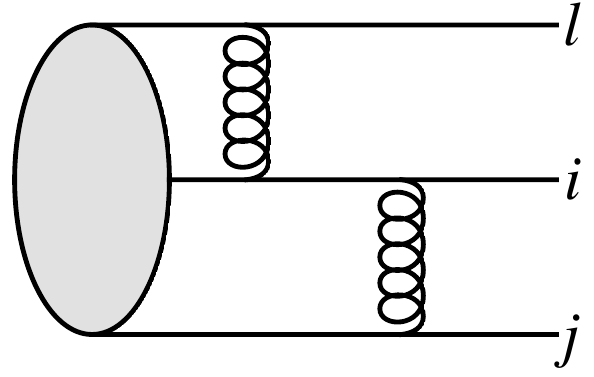}
\end{gathered}
=-\alpha_s^2 \sum_{i,j,l} (\bold{T}_i \cdot \bold{T}_l)(\bold{T}_i\cdot \bold{T}_l) (p_i\cdot p_l)(p_i\cdot p_j)\,  I_1^{(ijl)} \ ,
\end{split}
\end{equation}
\begin{equation}
\begin{split}
\begin{gathered}
\vspace{-0.2cm}
\includegraphics[width=2.2cm]{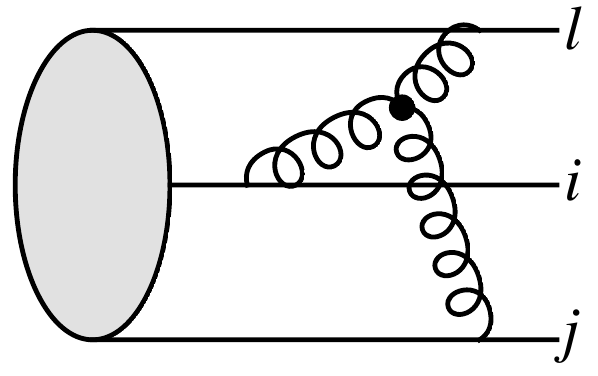}
\end{gathered}
=\frac{\alpha_s^2}{2} \sum_{i,j,l} i f^{abc} \bold{T}_i^a \bold{T}_j^b \bold{T}_l^c \, 2\left[(p_j\cdot p_l) I_2^{(ijl)}-(p_i\cdot p_j) I_3^{(ijl)}- (p_i\cdot p_l) I_4^{(ijl)}\right] \ .
\end{split}
\end{equation}
The two-loop integrals in the equations above are given by
\begin{equation}
\begin{aligned}
&I^{(ijl)}_1=\mu^{4\epsilon} \int_{k,q} \frac{1}{[p_i\cdot(k+q)+i0][p_i\cdot q+i0][k^2+i0][q^2+i0][-p_j\cdot k+i0][-p_l\cdot q+i0]} \\
&I^{(ijl)}_2= \mu^{4\epsilon} \int_{k,q} \frac{p_i\cdot k}{[-p_i\cdot q+i0][k^2+i0][q^2+i0][(k+q)^2+i0][-p_j\cdot k+i0][p_l\cdot (k+q)+i0]}\\
&I^{(ijl)}_3=\mu^{4\epsilon} \int_{k,q} \frac{p_l\cdot k}{[-p_i\cdot q+i0][k^2+i0][q^2+i0][(k+q)^2+i0][-p_j\cdot k+i0][p_l\cdot (k+q)+i0]} \\
&I^{(ijl)}_4=\mu^{4\epsilon} \int_{k,q} \frac{p_j\cdot q}{[-p_i\cdot q+i0][k^2+i0][q^2+i0][(k+q)^2+i0][-p_j\cdot k+i0][p_l\cdot (k+q)+i0]} \ ,
 \end{aligned}
\label{2ldiagram_3lines1}
\end{equation}
where the non-abelian three-line diagram gives an irreducible contribution. In this case the denominator structure is the same for $I^{(ijl)}_2, I^{(ijl)}_3$ and $I^{(ijl)}_4$, however, the numerator changes. This has to be taken into account when the shifts of the loop momenta are performed in the procedure of the Feynman tree theorem. In the following the FTT is applied to the integrals in Eq. (\ref{2ldiagram_3lines1}):
\vspace{-2cm}
\begin{equation}
\begin{split}
I^{(ijl)}_1= \mu^{4\epsilon} \int_{k,q} &\left\{ \frac{(2\pi i)^2 \tilde{\delta}(k) \tilde{\delta}(q)}{[p_i\cdot (k+q)+i0][p_i\cdot q+i0][-p_j\cdot k+i0][-p_l\cdot q+i0]} \right. \\
&\left. +\frac{(2\pi i)^2 \tilde{\delta}(k) \delta(p_i\cdot q)}{[p_i\cdot (k+q)+i0][q^2+i0][-p_j\cdot k+i0][-p_l\cdot q+i0]}\right. \\
&\left. +\frac{(2\pi i)^2 \tilde{\delta}(q) \delta(p_i\cdot k)}{[p_i\cdot q+i0][(k-q)^2+i0][p_j\cdot (q-k)+i0][-p_l\cdot q+i0]} \right. \\
&\left. + \frac{(2\pi i)^2 \delta(p_i\cdot k)\delta(p_j\cdot q)}{[p_i\cdot(k+q)+i0][(k+q)^2+i0][p_j\cdot q+i0][-p_l\cdot (k+q)+i0]}\right. \\
& \left. +\frac{(2\pi i)^2 \delta(p_i\cdot k) \delta(p_i\cdot q)}{[q^2+i0][(k-q)^2+i0][p_j\cdot (q-k)+i0][-p_l\cdot q+i0]}\right. \\
&\left. - \frac{(2\pi i)^2 \tilde{\delta(k)} \delta(p_i\cdot q)}{[p_i\cdot (q-k)+i0][(q-k)^2+i0][-p_j\cdot k+i0][p_l\cdot (k-q)+i0]}\right. \\
& \left. + \frac{(2\pi i)^3 \tilde{\delta}(k-q) \delta(p_i\cdot k) \delta(p_i\cdot q)}{[q^2+i0][p_j\cdot (q-k)+i0][-p_l\cdot q+i0]}\right. \\
&\left. + \frac{(2\pi i)^3 \tilde{\delta}(k-q) \delta(p_i\cdot k) \delta(p_j\cdot (q-k))}{[p_i\cdot q+i0][q^2+i0][-p_l\cdot q+i0]}\right. \\
& \left. + \frac{(2\pi i)^3 \tilde{\delta}(q) \delta(p_i\cdot k) \delta(p_j\cdot (q-k))}{[p_i\cdot q+i0][(k-q)^2+i0][-p_l\cdot q+i0]}\right. \\
&\left. + \frac{(2\pi i)^3 \delta(p_i\cdot k) \delta(p_i\cdot q) \delta(p_j\cdot(q-k))}{[q^2+i0][(k-q)^2+i0][-p_l\cdot q+i0]}\right. \\
&\left. +\frac{(2\pi i)^3 \tilde{\delta}(q) \tilde{\delta}(k-q) \delta(p_i\cdot k)}{[p_i\cdot q+i0][p_j\cdot (q-k)+i0][-p_l\cdot q+i0]}\right. \\
&\left. +\frac{(2\pi i)^4 \tilde{\delta}(k-q) \delta(p_i\cdot q) \delta(p_i\cdot k) \delta(p_j\cdot(q-k))}{[q^2 +i0][-p_l\cdot q+i0]}\right\} \ .
\end{split}
\label{2ldiagram_xiii}
\end{equation}
\begin{equation}
\begin{split}
I^{(ijl)}_2=\mu^{4\epsilon} \int_{k,q} & \left\{ - \frac{(2\pi i)^2 (p_i\cdot k) \tilde{\delta}(q) \delta(p_l\cdot k)}{[-p_i\cdot q+i0][k^2+i0][(k-q)^2+i0][p_j\cdot (q-k)+i0]}\right. \\
& \left. - \frac{(2\pi i)^2 (p_i\cdot k) \tilde{\delta}(k) \tilde{\delta}(q)}{[-p_i\cdot q+i0][(k-q)^2+i0][p_j\cdot (q-k)+i0][p_l\cdot k+i0]} \right. \\
&\left. +\frac{(2\pi i)^2 (p_i\cdot k) \tilde{\delta}(q) \tilde{\delta}(k)}{[-p_i\cdot q+i0][(k+q)^2+i0][-p_j\cdot k+i0][p_l\cdot (k+q)+i0]} \right. \\
&\left. +\frac{(2\pi i)^2 (p_i\cdot k) \tilde{\delta}(q) \tilde{\delta}(k)}{[p_i\cdot (k-q)+i0][(q-k)^2+i0][-p_j\cdot k+i0][p_l\cdot q+i0]} \right. \\
&\left. +\frac{(2\pi i)^2 (p_i\cdot q) \tilde{\delta}(q) \delta(p_i\cdot k)}{[k^2+i0][(k+q)^2+i0][-p_j\cdot (k+q)+i0][p_l\cdot q+i0]}\right. \\
&\left. +\frac{(2\pi i)^2 (p_i\cdot k) \tilde{\delta}(k) \delta(p_l\cdot q)}{[p_i\cdot (k-q)+i0][q^2+i0][(q-k)^2+i0][-p_j\cdot k+i0]}\right. \\
&\left. + \frac{(2\pi i)^2 (p_i\cdot q) \delta(p_i\cdot k) \delta(p_l\cdot q)}{[q ^2+i0][k^2+i0][(k+q)^2+i0][-p_j\cdot (k+q)+i0]} \right. \\
& \left. +\frac{(2\pi i)^3 (p_i\cdot k) \tilde{\delta}(q) \tilde{\delta}(k) \tilde{\delta}(k+q)}{[-p_i\cdot q+i0][-p_j\cdot k+i0][p_l\cdot(k+q)+i0]}\right. \\
& \left. +\frac{(2\pi i)^3 (p_i\cdot k)\tilde{\delta}(q) \tilde{\delta}(k) \delta(p_l\cdot (k+q))}{[-p_i\cdot q+i0][(k+q)^2+i0][-p_j\cdot k+i0]}\right. \\
&\left. + \frac{(2\pi i)^3 (p_i\cdot k) \tilde{\delta}(q) \tilde{\delta}(k) \tilde{\delta}(q-k)}{[p_i\cdot (k-q)+i0][-p_j\cdot k+i0][p_l\cdot q+i0]}\right. \\
& \left. + \frac{(2\pi i)^3 (p_i\cdot k) \tilde{\delta}(k) \tilde{\delta}(q) \delta(p_i\cdot (k-q))}{[(q-k)^2+i0][-p_j\cdot k+i0][p_l\cdot q+i0]}\right. \\
&\left. + \frac{(2\pi i)^3 (p_i\cdot k) \tilde{\delta}(q) \tilde{\delta}(q-k) \delta(p_i\cdot (k-q))}{[k^2+i0][-p_j\cdot k+i0][p_l\cdot q+i0]}\right. \\
&\left. + \frac{(2\pi i)^3 (p_i\cdot k) \tilde{\delta}(q-k) \delta(p_l\cdot q) \delta(p_i\cdot(k-q))}{[q^2+i0][k^2+i0][-p_j\cdot k+i0]}\right. \\
&\left. + \frac{(2\pi i)^3 (p_i\cdot k) \tilde{\delta}(k) \delta(p_l\cdot q) \delta(p_i\cdot (k-q))}{[(q-k)^2+i0][q^2+i0][-p_j\cdot k+i0]}\right. \\
&\left. + \frac{(2\pi i)^3 (p_i\cdot k) \tilde{\delta}(k) \tilde{\delta}(q-k) \delta(p_l\cdot q)}{[p_i\cdot (k-q)+i0][q^2+i0][-p_j\cdot k+i0]}\right\} \ , 
\end{split}
\label{2ldiagram_xiv}
\end{equation}
\begin{equation}
\begin{split}
I^{(ijl)}_3=\mu^{4\epsilon} \int_{k,q} & \left\{ \frac{(2\pi i)^2 (p_l\cdot q) \tilde{\delta}(q) \delta(p_l\cdot k)}{[-p_i\cdot q+i0][k^2+i0][(k-q)^2+i0][p_j\cdot (q-k)+i0]}\right. \\
& \left. + \frac{(2\pi i)^2 (p_l\cdot q) \tilde{\delta}(k) \tilde{\delta}(q)}{[-p_i\cdot q+i0][(k-q)^2+i0][p_j\cdot (q-k)+i0][p_l\cdot k+i0]} \right. \\
&\left. +\frac{(2\pi i)^2 (p_l\cdot k) \tilde{\delta}(q) \tilde{\delta}(k)}{[-p_i\cdot q+i0][(k+q)^2+i0][-p_j\cdot k+i0][p_l\cdot (k+q)+i0]} \right. \\
&\left. +\frac{(2\pi i)^2 (p_l\cdot k) \tilde{\delta}(q) \tilde{\delta}(k)}{[p_i\cdot (k-q)+i0][(q-k)^2+i0][-p_j\cdot k+i0][p_l\cdot q+i0]} \right. \\
&\left. +\frac{(2\pi i)^2 (p_l\cdot k) \tilde{\delta}(q) \delta(p_i\cdot k)}{[k^2+i0][(k+q)^2+i0][-p_j\cdot (k+q)+i0][p_l\cdot q+i0]}\right. \\
&\left. +\frac{(2\pi i)^2 (p_l\cdot k) \tilde{\delta}(k) \delta(p_l\cdot q)}{[p_i\cdot (k-q)+i0][q^2+i0][(q-k)^2+i0][-p_j\cdot k+i0]}\right. \\
&\left. + \frac{(2\pi i)^2 (p_l\cdot k) \delta(p_i\cdot k) \delta(p_l\cdot q)}{[q ^2+i0][k^2+i0][(k+q)^2+i0][-p_j\cdot (k+q)+i0]} \right. \\
& \left. +\frac{(2\pi i)^3 (p_l\cdot k) \tilde{\delta}(q) \tilde{\delta}(k) \tilde{\delta}(k+q)}{[-p_i\cdot q+i0][-p_j\cdot k+i0][p_l\cdot(k+q)+i0]}\right. \\
& \left. +\frac{(2\pi i)^3 (p_l\cdot k) \tilde{\delta}(q) \tilde{\delta}(k) \delta(p_l\cdot (k+q))}{[-p_i\cdot q+i0][(k+q)^2+i0][-p_j\cdot k+i0]}\right. \\
&\left. + \frac{(2\pi i)^3 (p_l\cdot k) \tilde{\delta}(q) \tilde{\delta}(k) \tilde{\delta}(q-k)}{[p_i\cdot (k-q)+i0][-p_j\cdot k+i0][p_l\cdot q+i0]}\right. \\
& \left. + \frac{(2\pi i)^3 (p_l\cdot k) \tilde{\delta}(k) \tilde{\delta}(q) \delta(p_i\cdot (k-q))}{[(q-k)^2+i0][-p_j\cdot k+i0][p_l\cdot q+i0]}\right. \\
&\left. + \frac{(2\pi i)^3 (p_l\cdot k) \tilde{\delta}(q) \tilde{\delta}(q-k) \delta(p_i\cdot (k-q))}{[k^2+i0][-p_j\cdot k+i0][p_l\cdot q+i0]}\right. \\
&\left. + \frac{(2\pi i)^3 (p_l\cdot k) \tilde{\delta}(q-k) \delta(p_l\cdot q) \delta(p_i\cdot(k-q))}{[q^2+i0][k^2+i0][-p_j\cdot k+i0]}\right. \\
&\left. + \frac{(2\pi i)^3 (p_l\cdot k) \tilde{\delta}(k) \delta(p_l\cdot q) \delta(p_i\cdot (k-q))}{[(q-k)^2+i0][q^2+i0][-p_j\cdot k+i0]}\right. \\
&\left. + \frac{(2\pi i)^3 (p_l\cdot k) \tilde{\delta}(k) \tilde{\delta}(q-k) \delta(p_l\cdot q)}{[p_i\cdot (k-q)+i0][q^2+i0][-p_j\cdot k+i0]}\right\} \ , 
\end{split}
\label{2ldiagram_xv}
\end{equation}
\begin{equation}
\begin{split}
I^{(ijl)}_4=\mu^{4\epsilon} \int_{k,q} & \left\{- \frac{(2\pi i)^2 (p_j \cdot q) \tilde{\delta}(q) \delta(p_l\cdot k)}{[-p_i\cdot q+i0][k^2+i0][(k-q)^2+i0][p_j\cdot (q-k)+i0]}\right. \\
& \left. - \frac{(2\pi i)^2 (p_j\cdot q) \tilde{\delta}(k) \tilde{\delta}(q)}{[-p_i\cdot q+i0][(k-q)^2+i0][p_j\cdot (q-k)+i0][p_l\cdot k+i0]} \right. \\
&\left. + \frac{(2\pi i)^2 (p_j\cdot q) \tilde{\delta}(q) \tilde{\delta}(k)}{[-p_i\cdot q+i0][(k+q)^2+i0][-p_j\cdot k+i0][p_l\cdot (k+q)+i0]} \right. \\
&\left. + \frac{(2\pi i)^2 (p_j\cdot q) \tilde{\delta}(q) \tilde{\delta}(k)}{[p_i\cdot (k-q)+i0][(q-k)^2+i0][-p_j\cdot k+i0][p_l\cdot q+i0]} \right. \\
&\left. - \frac{(2\pi i)^2 (p_j\cdot k) \tilde{\delta}(q) \delta(p_i\cdot k)}{[k^2+i0][(k+q)^2+i0][-p_j\cdot (k+q)+i0][p_l\cdot q+i0]}\right. \\
&\left. + \frac{(2\pi i)^2 (p_j\cdot q) \tilde{\delta}(k) \delta(p_l\cdot q)}{[p_i\cdot (k-q)+i0][q^2+i0][(q-k)^2+i0][-p_j\cdot k+i0]}\right. \\
&\left. - \frac{(2\pi i)^2 (p_j\cdot k) \delta(p_i\cdot k) \delta(p_l\cdot q)}{[q ^2+i0][k^2+i0][(k+q)^2+i0][-p_j\cdot (k+q)+i0]} \right. \\
&\left. + \frac{(2\pi i)^3 (p_j\cdot q) \tilde{\delta}(q) \tilde{\delta}(k) \tilde{\delta}(k+q)}{[-p_i\cdot q+i0][-p_j\cdot k+i0][p_l\cdot(k+q)+i0]}\right. \\
&\left. + \frac{(2\pi i)^3 (p_j\cdot q) \tilde{\delta}(q) \tilde{\delta}(k) \delta(p_l\cdot (k+q))}{[-p_i\cdot q+i0][(k+q)^2+i0][-p_j\cdot k+i0]}\right. \\
&\left. + \frac{(2\pi i)^3 (p_j\cdot q) \tilde{\delta}(q) \tilde{\delta}(k) \tilde{\delta}(q-k)}{[p_i\cdot (k-q)+i0][-p_j\cdot k+i0][p_l\cdot q+i0]}\right. \\
&\left. + \frac{(2\pi i)^3 (p_j\cdot q) \tilde{\delta}(k) \tilde{\delta}(q) \delta(p_i\cdot (k-q))}{[(q-k)^2+i0][-p_j\cdot k+i0][p_l\cdot q+i0]}\right. \\
&\left. + \frac{(2\pi i)^3 (p_j\cdot q) \tilde{\delta}(q) \tilde{\delta}(q-k) \delta(p_i\cdot (k-q))}{[k^2+i0][-p_j\cdot k+i0][p_l\cdot q+i0]}\right. \\
&\left. + \frac{(2\pi i)^3 (p_j\cdot q) \tilde{\delta}(q-k) \delta(p_l\cdot q) \delta(p_i\cdot(k-q))}{[q^2+i0][k^2+i0][-p_j\cdot k+i0]}\right. \\
&\left. + \frac{(2\pi i)^3 (p_j\cdot q) \tilde{\delta}(k) \delta(p_l\cdot q) \delta(p_i\cdot (k-q))}{[(q-k)^2+i0][q^2+i0][-p_j\cdot k+i0]} \right. \\
&\left. + \frac{(2\pi i)^3 (p_j\cdot q) \tilde{\delta}(k) \tilde{\delta}(q-k) \delta(p_l\cdot q)}{[p_i\cdot (k-q)+i0][q^2+i0][-p_j\cdot k+i0]}\right\} \ . 
\end{split}
\label{2ldiagram_xvi}
\end{equation}

\bibliography{nlo-colorflow-evolution}

\end{document}